# An Exact General-Relativity Solution for the Motion and Intersections of Self-Gravitating Shells in the Field of a Massive Black Hole

M. V. Barkov[a], *, V. A. Belinski[b], **, and G. S. Bisnovatyi-Kogan[c], ***

[a]Space Research Institute, Russian Academy of Sciences, ul. Profsoyuznaya 84/32, Moscow, 117810 Russia
*e-mail: barmv@sai.msu.ru
[b]National Institute of Nuclear Research (INFN) and International Center of Relativistic Astrophysics (ICRA), Rome, Italy
**e-mail: belinski@icra.it
[c]Institute des Hautes Études Scientifiques (IHES), Bures-sur-Yvette, France
***e-mail: gkogan@mx.iki.rssi.ru

March 13, 2002

**Abstract**—The motion with intersections of relativistic gravitating shells in the Schwarzschild gravitational field of a central body is considered. Formulas are derived for calculating parameters of the shells after intersection via their parameters before intersection. Such special cases as the Newtonian approximation, intersections of light shells, and intersections of a test shell with a gravitating shell are also considered. The ejection of one of the shells to infinity in the relativistic region is described. The equations of motion for the shells are analyzed numerically.

## 1. INTRODUCTION

An interesting application of the theory of gravitating, spherically symmetric shells is related to the possibility of modeling some properties of star clusters by the evolution of such shells, each moving in the field produced by all the remaining shells and the central body [1–6]. Let there be steady motion of a large number of gravitating particles (stars) around a nonrotating central body at rest. Because of the large number of particles, each of them may be assumed to move in an average stationary spherically symmetric field (which is not a Schwarzschild field inside the cluster but transforms into a Schwarzschild field outside the cluster). Each particle moves along a trajectory that is a generalization of elliptical orbits in Newtonian mechanics; the total energy and total angular momentum of each particle are conserved. Let us single out a sphere of arbitrary radius $r_0$ inside the cluster at some arbitrary time $t_0$ and single out the particles on this sphere with equal radial velocities and with the same absolute value of the total angular momentum per unit rest mass, whence it follows that the total energies per unit rest mass are also equal. One can see from the first integrals of geodesics in a stationary spherically symmetric field that the radial motion of all the singled-out particles obeys the same equations. This implies that the radial motion of all these particles is the same; i.e., they all form what is called a shell. Whereas the shell as a whole moves only radially, its constituent particles also move tangentially inside the shell. When the number of particles is large, the tangential particle motions may be treated as chaotic motions with the parameters distributed uniformly and isotropically on the two-dimensional shell surface. Consequently, these motions can be taken into account hydrodynamically as the presence of a tangential pressure. The radial behavior of a cluster is modeled by the motion of a discrete set of infinitely thin shells, each moving in a vacuum. The radial motion is accurately taken into account by solving the gravitational equations. Of course, not all cluster properties can be described in terms of this model but several aspects of its dynamics are well modeled. These include, for example, the violent relaxation of a star cluster [1, 5] and the ejection of some of the cluster layers during its collapse after the loss of stability.

Clearly, the model described above has the following two main determining elements: first, the motion of one shell in the field of a central mass and, second, the motion of two shells in the field of a central mass, including the possible intersections of the shells with one another. The first problem was solved by Chase [7] more than thirty years ago. This author did not raise the question of applications to analysis of the cluster behavior but derived the equation of motion for one shell with a tangential pressure (in general, also with an electric charge) and studied its stability against collapse. He derived the equation of motion for a shell in a general form, i.e., without making any special assumptions about the equation of state for the shell matter.

Up until now, there has been no complete general-relativity solution to the second problem. We hope that this paper will bridge this gap to some extent. We emphasize that, unless the shells intersect, their equations of motion are a trivial generalization of the equa-tion of motion for one shell. This is because a spherically symmetric shell produces no gravitational field within itself and, therefore, does not affect anything inside it. At the same time, the effect of the inner shell on the outer shell can be easily taken into account: only the mass parameters of the Schwarzschild metrics in outer (relative to the inner shell) regions change. Thus, the main problem here is related precisely to the intersection of shells and involves allowance for the effects of a large number of such intersections. In this case, we treat the intersection as a purely ballistic process; i.e., each shell is affected only by its surrounding gravitational field, and there is no direct nongravitational interaction between the shells at the time of their intersection. For this condition to be satisfied, the joining must be such that, in the limit of test shells (whose gravitation may be ignored), they would intersect without responding in any way to this process; i.e., all parameters of the motion of each shell in this limit would remain continuous when passing through the point of intersection.

Note that the above problem was considered in terms of Newtonian gravitation in [1–6, 8]. An interesting result (first obtained in [4]) is the ejection of one of the shells to spatial infinity through energy transfer between the shells as they intersect. In [8], we also described this effect but with some new features. In addition, we numerically solved the equations of motion for two shells by taking into account a large number of their intersections with one another. We found that, after a sufficiently large number of intersections, the motion of the shells became chaotic with the strong sensitivity to small variations in initial parameters typical of chaos.

Noteworthy are also [9, 10], in which the motion and intersection of shells was considered in a strong gravitational field. As our comparison with the exact solution presented here shows, the approximate procedure used in these papers to describe the intersection has a high accuracy only when one (or both) of the velocities of the intersecting shells is low and when the ratios of the effective shell masses to the central-body mass are small.

Finally, we emphasize that our study is intended for specific astrophysical applications and, of course, it does not cover all aspects of the theory for gravitating shells. Therefore, our list of references is quite limited and reflects the development of those applied issues that are dealt with in this paper. Among the most recent papers that have a direct bearing on our study, the paper by Berezin and Okhrimenko [11] is particularly noteworthy. These authors also investigated the behavior of a gravitating shell composed of particles moving in the field of a central mass and considered both astrophysical and quantum aspects of the problem. However, the authors did not touch on the issue of the intersection of two shells. Neronov [12] considered the problem of inelastic collisions between an arbitrary number of $n$-dimensional shells in $(n + 2)$-dimensional space-time and pointed out the conservation law that related a certain combination of shell parameters before collision with a similar combination of shell parameters after collision. The author applied this result to the collision of three-dimensional shells in five-dimensional anti–de Sitter space-time to study the properties of the so-called ekpyrotyc cosmological model [13, 14]. For completeness, note also [15], which mainly repeats Neronov's results. The interested reader will find in the cited papers [11–15] quite an extensive review of the literature on various aspects of the theory for thin gravitating shells.

## 2. A GRAVITATING SHELL WITH A TANGENTIAL PRESSURE

Chase [7] obtained his result by the geometrical method that was first used in [16], where the equation of motion for a spherically symmetric dust shell was derived by this method. Naturally, the same result can be obtained in a more habitual (for the physicist) way by making up the energy–momentum tensor with an appropriate δ-shaped source and by directly integrating the Einstein equations with this right-hand side. We used precisely this derivation method, and our result closely matched Chase's result. Of course, we do not detail our calculations here but only note the main points, because this is of methodological interest and, in addition, provides a convenient means of introducing all the necessary concepts and notation.

Let there be a central body of mass $m_{in}$ and let a spherically symmetric shell move outside this body. Even before solving any equations, it is clear that the metric inside and outside the shell is the Schwarzschild metric but with different mass parameters. Using the coordinates $x^0 = ct$ and $r$, which are continuous when passing through the shell, we can write the intervals[1] inside, outside, and on the shell as

$$-(ds^2)_{in} = -e^{T(t)} f_{in}(r) c^2 dt^2 + f_{in}^{-1}(r) dr^2 + r^2 d\Omega^2, \quad (1)$$

$$-(ds^2)_{out} = -f_{out}(r) c^2 dt^2 + f_{out}^{-1}(r) dr^2 + r^2 d\Omega^2, \quad (2)$$

$$-(ds^2)_{on} = -c^2 d\tau^2 + r_0^2(\tau) d\Omega^2, \quad (3)$$

where we denoted

$$d\Omega^2 = d\theta^2 + \sin^2\theta d\phi^2$$

and

$$f_{in} = 1 - \frac{2km_{in}}{c^2 r}, \quad f_{out} = 1 - \frac{2km_{out}}{c^2 r}. \quad (4)$$

In the interval (3), $\tau$ is the proper time of the shell. The factor $e^T$ in (1) is required to ensure that the time coordinate $t$ be continuous when passing through the shell. The parameter $m_{out} c^2$ is the total energy of the system.

---

[1] The interval is written as $-ds^2 = g_{ik} dx^i dx^k$ and the metric signature is $(-, +, +, +)$, i.e., $g_{00} < 0$. The Roman indices take on 0, 1, 2, and 3. The standard notation for spherical coordinates is $(x^0, x^1, x^2, x^3) = (ct, r, \theta, \phi)$. The Newtonian gravitational constant is denoted by $k$.

If the equation of motion for the shell is

$$r = R_0(t),$$

then joining the angular parts of all three intervals (1)–(3) yields

$$r_0(\tau) = R_0[t(\tau)], \quad (5)$$

where the function $t(\tau)$ describes the relationship between the global time and the proper time of the shell. Joining the radial-time parts of the intervals (1)–(3) on the shell requires that the following relations hold:

$$f_{\text{in}}(r_0)\left(\frac{dt}{d\tau}\right)^2 e^{T(t)} - f_{\text{in}}^{-1}(r_0)\left(\frac{dr_0}{cd\tau}\right)^2 = 1, \quad (6)$$

$$f_{\text{out}}(r_0)\left(\frac{dt}{d\tau}\right)^2 - f_{\text{out}}^{-1}(r_0)\left(\frac{dr_0}{cd\tau}\right)^2 = 1. \quad (7)$$

If the equation of motion for the shell [i.e., the function $r_0(\tau)$] is known, then the function $t(\tau)$ follows from (7) and we can then derive $T(t)$ from (6). Thus, the problem consists only in determining $r_0(\tau)$, which can be done by directly integrating the Einstein equations for the metric

$$-(ds^2) = g_{00}(t, r)c^2 dt^2 + g_{11}(t, r)dr^2 + r^2 d\Omega^2 \quad (8)$$

and the energy–momentum tensor

$$T_i^k = \varepsilon u_i u^k + (\delta_i^2 \delta_2^k + \delta_i^3 \delta_3^k) p. \quad (9)$$

Here, $u_2 = u_3 = 0$ and $\varepsilon$, $p$, $u_0$, and $u_1$ depend on the coordinates $t$ and $r$ alone, with

$$u^0 u_0 + u^1 u_1 = -1.$$

The energy density $\varepsilon$ is

$$\varepsilon = \frac{M(t)c^2 \delta[r - R_0(t)]}{4\pi r^2 u^0 \sqrt{-g_{00} g_{11}}}, \quad (10)$$

where $\delta$ is the standard $\delta$ function. In the absence of tangential pressure $p$, the quantity $M$ in (10) would be constant and the energy density $\varepsilon$ would be the sum of the rest energies of the shell particles per unit volume of the radially comoving frame. In this case, $M$ would be the total rest mass of the shell or its bare or baryonic mass. In the presence of pressure, $Mc^2$ includes the rest energy along with the energy (in the radially comoving frame) of the tangential motions of the shell particles that produce this pressure. In this case, $Mc^2$ can no longer be constant but depends on the degree of compression of the shell, i.e., on its radius $R_0(t)$ or simply on time $t$, which is explicitly specified in (10).

As in any spherically symmetric problem, we can choose the hydrodynamic equations $T^k_{i;k} = 0$ and the $\binom{0}{0}$, $\binom{1}{1}$, and $\binom{1}{0}$ components of the Einstein equations taken in the form

$$R_i^k - \frac{1}{2}\delta_i^k R = \frac{8\pi k}{c^4} T_i^k$$

(all the remaining components of these equations are identically satisfied either in view of the Bianchi identities or because of the symmetry of the problem) as the complete system of equations. As we see from (9), these components of the Einstein equations contain no pressure. It is easy to show that they actually lead[2] to the solution (1)–(4) with arbitrary constants $m_{\text{in}}$ and $m_{\text{out}}$ and, in addition, to the following first-order equation for the function $r_0(\tau)$:

$$\sqrt{f_{\text{in}}(r_0) + \left(\frac{dr_0}{cd\tau}\right)^2} + \sqrt{f_{\text{out}}(r_0) + \left(\frac{dr_0}{cd\tau}\right)^2} = \frac{2(m_{\text{out}} - m_{\text{in}})}{\mu(\tau)}, \quad (11)$$

where we denoted

$$\mu(\tau) = M[t(\tau)]. \quad (12)$$

Two of the four equations $T^k_{i;k} = 0$ are identically satisfied because of the symmetry of the problem. The other two equations can be reduced to such a form that one of them will serve simply to determine the pressure,

$$p = -\frac{dM}{dt} \frac{c\delta[r - R_0(t)]}{8\pi r u^1 \sqrt{-g_{00} g_{11}}}, \quad (13)$$

and the other is identically satisfied after substituting this expression for $p$ in it (it is a differential equation of the second order in $\tau$ for the function $r_0(\tau)$, which is nothing else but the result of differentiating Eq. (11) with respect to $\tau$). It should be noted, however, that the latter holds only in the general unsteady-state case where $R_0 \neq \text{const}$.

The case of a steady-state shell, $R_0 = \text{const}$, requires a special analysis. In this special case, all quantities depend only on one variable $r$ and $u^1 = 0$. Three of the four equations $T^k_{i;k} = 0$ are identically satisfied and the

---

[2] When integrating the equations of the problem under consideration, we need only two standard rules to work with symbolic functions: $d\theta(x)/dx = \delta(x)$ (where $\theta$ is the Heaviside step function) and $F(x)\delta(x) = (1/2)[F(-0) + F(+0)]\delta(x)$.

fourth equation again serves to determine the pressure, which can be written as

$$p = \frac{Mc^2}{32\pi R_0^2} \times \left[\frac{1 - f_{in}(R_0)}{\sqrt{f_{in}(R_0)}} + \frac{1 - f_{out}(R_0)}{\sqrt{f_{out}(R_0)}}\right]\delta(r - R_0). \quad (14)$$

Expression (10) for the energy density now takes the form

$$\varepsilon = \frac{Mc^2}{8\pi R_0^2}[\sqrt{f_{in}(R_0)} + \sqrt{f_{out}(R_0)}]\delta(r - R_0). \quad (15)$$

The shell radius $R_0$ itself must be determined from Eq. (11), in which we should now set $dr_0/d\tau = 0$ and $\mu = M =$ const. It is easy to establish that a positive solution for $R_0$ at positive parameters $m_{in}$ and $m_{out}$ exists only when the inequality

$$(m_{out} - m_{in})^2 < M^2 \quad (16)$$

is satisfied; this solution is

$$R_0 = \frac{kM^2}{2c^2}\frac{m_{out} + m_{in} + \sqrt{4m_{out}m_{in} + M^2}}{M^2 - (m_{out} - m_{in})^2}. \quad (17)$$

Expressions (14)–(17) give the solution to the problem for a steady-state shell.

Let us return to the general unsteady-state case. The equation of motion for the shell (11) can be written in several equivalent forms. Below, we give the following two forms:

$$\sqrt{f_{in}(r_0) + \left(\frac{dr_0}{cd\tau}\right)^2} = \frac{m_{out} - m_{in}}{\mu(\tau)} + \frac{k\mu(\tau)}{2c^2 r_0}, \quad (18)$$

$$\sqrt{f_{out}(r_0) + \left(\frac{dr_0}{cd\tau}\right)^2} = \frac{m_{out} - m_{in}}{\mu(\tau)} - \frac{k\mu(\tau)}{2c^2 r_0}. \quad (19)$$

Given expressions (4) for $f_{in}$ and $f_{out}$, it is easy to verify that both (18) and (19) directly follow from Eq. (11). At the same time, Eq. (11) is the sum of Eqs. (18) and (19). Yet another equivalent form of the equations of motion for the shell can be derived by squaring each of Eqs. (18) and (19) and then adding them term by term. As a result, we obtain

$$1 + \left(\frac{dr_0}{cd\tau}\right)^2 = \frac{(m_{out} - m_{in})^2}{\mu^2(\tau)} + \frac{k(m_{out} + m_{in})}{c^2 r_0} + \frac{k^2\mu^2(\tau)}{4c^4 r_0^2}. \quad (20)$$

The equation of motion for the shell is given in [7] precisely in this form. We emphasize that, for actual astrophysical applications, all the radicals encountered in our paper should be taken to be positive.

To proceed further, we must specify the function $\mu(\tau)$, which, as we see from (10) and (13), is equivalent to specifying an equation of state. Here, of course, there is a wide range of possibilities to choose from, but we restrict our analysis to the shell model described in the Introduction, i.e., a shell composed of particles that move in the field of a central mass. The quantity $Mc^2$ in (10) in the absence of pressure is the total rest energy of the shell, i.e.,

$$Mc^2 = \sum_a m_a c^2,$$

where the sum is taken over all particles and $m_a$ is the rest mass of the individual particle. The shell is meant to be in the state of rest with respect to its radial motion, and this state takes place in the radially comoving frame. Since the tangential motions are normal to the radial motion, their role reduces only to producing an effective rest mass with respect to the radial motion of the entire shell; i.e., in their presence, we have the following expression for $Mc^2$:

$$Mc^2 = \sum_a \sqrt{m_a^2 c^4 + p_a^2 c^2} = \sum_a \left(m_a c^2 \sqrt{1 + \frac{p_a^2}{m_a^2 c^2}}\right), \quad (21)$$

where $p_a$ is the tangential momentum of each particle. By the definition of the shell (see the Introduction), all its particles are at the same distance $R_0$ from the center and they all have the same $|l_a|/m_a$ ratio (where $l_a$ is the total angular momentum of particle $a$). Since

$$\frac{p_a^2}{m_a^2} = \frac{l_a^2}{m_a^2 R_0^2} = \frac{\text{const}}{R_0}, \quad (22)$$

the square root in (21) does not depend on the index $a$ and this root can be taken outside the sum and the sum

$$\sum_a m_a c^2 = mc^2,$$

where the constant $m$ is the total rest mass of the entire shell. Because the $|l_a|/m_a$ ratio is independent of $a$, we have

$$\left(\sum_a m_a c^2\right)^2 \frac{p_a^2}{m_a^2 c^2} = \frac{c^2}{R_0^2}\left(\frac{|l_a|}{m_a}\sum_a m_a\right)^2 = \frac{c^2}{R_0^2}\left(\sum_a |l_a|\right)^2. \quad (23)$$

As a result, formula (21) [given the designation (12)] can be written as

$$\mu(\tau) = \sqrt{m^2 + \frac{L^2}{c^2 r_0^2(\tau)}}, \quad (24)$$

where

$$L = \sum_a |l_a|$$

is the sum of the absolute values of the total angular momenta for the shell particles. Substituting (24) in Eq. (11) [or in one of the equivalent equations (18)–(20)] yields the final equation. We can determine the function $r_0(\tau)$ from this equation if the initial shell radius and the four arbitrary constants $m_{\text{in}}$, $m_{\text{out}}$, $m$, and $L$ are specified.

According to (10), (12), (13), and (24), the equation of state that relates the shell energy density $\varepsilon$ to the tangential pressure $p$ is

$$p = \frac{\varepsilon}{2} \frac{L^2}{m^2 c^2 R_0^2} \left(1 + \frac{L^2}{m^2 c^2 R_0^2}\right)^{-1}.$$

The relation

$$\frac{L^2}{m^2 c^2 R_0^2} = \frac{p_a^2}{m_a^2 c^2}$$

(recall that $p_a^2/m_a^2$ do not depend on particle number $a$) follows from the definition of the constants $L$ and $m$ and from formula (23). Hence, we see the limiting forms of the equation of state: we have a dust state, $p \ll \varepsilon$, for nonrelativistic tangential velocities ($p_a^2 \ll m_a^2 c^2$) and obtain $p = (1/2)\varepsilon$ for ultrarelativistic tangential particle velocities ($p_a^2 \gg m_a^2 c^2$), as should be the case for a two-dimensional ultrarelativistic gas. As the shell expands to infinity ($R_0 \longrightarrow \infty$), the equation of state always tends to the dust one, because the contribution of the tangential particle motions becomes negligible.

## 3. THE INTERSECTION OF SHELLS

Let us now consider the next (in complexity) case of two shells that move in the field of a central mass. When the shells do not intersect, the equations of their motion can be immediately written without difficulty by using the previously derived equations of motion for one shell and the fact that a spherically symmetric shell has no effect on anything inside it. When there is an intersection, this information is no longer enough, because additional joining conditions that are not contained in the theory of motion for one shell are required at the point of intersection. As previously, we cover the part of the physical space-time of interest by a global coordinate system $(r, x^0) = (r, ct)$ with continuous $r$ and $t$. Let shell 1 be inside shell 2 at the initial time and in its vicinity and then let these shells intersect at some point of space-time $(r^*, t^*)$, so shell 2 turns out to be inside shell 1 after $t^*$ and these relative positions of the shells are maintained for some time after $t^*$. The intersection process is shown in Fig. 1. If $r = R_1(t)$ and $r = R_2(t)$ are

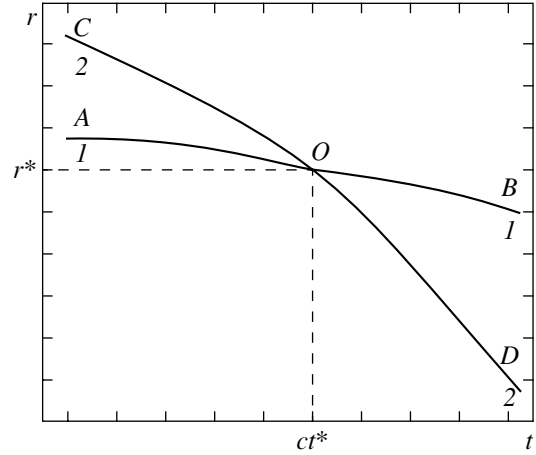

**Fig. 1.** A schematic view of the intersection of two gravitating shells. The coordinates of point $O$ are $ct^*$ and $r^*$.

the equations of motion for the first and second shells, then we have four space-time regions:

$$COB \quad (r > R_1, r > R_2),$$
$$COA \quad (R_1 < r < R_2),$$
$$AOD \quad (r < R_1, r < R_2),$$
$$BOD \quad (R_2 < r < R_1).$$

In each of these regions, the metric is the Schwarzschild metric but supplemented with the dilaton factor $e^{T(t)}$, which is required if we wish to cover all four regions by a global continuous time coordinate $t$. In this case, it remains possible to choose the metric coefficient $g_{00}$ in a purely Schwarzschild form without the dilaton factor in any of these four regions (but only in one of them), thereby fixing the choice of global time $t$. It would be natural to introduce a purely Schwarzschild metric outside the shells, i.e., in region $COB$ adjacent to spatial infinity. Thus, the metric in these four regions has the form (8) but with different metric coefficients $g_{00}$ and $g_{11}$:

$$g_{00}^{(COB)} = -f_{\text{out}}(r), \quad g_{11}^{(COB)} = f_{\text{out}}^{-1}(r), \quad (25)$$

$$g_{00}^{(COA)} = -e^{T_1(t)} f_{12}(r), \quad g_{11}^{(COA)} = f_{12}^{-1}(r), \quad (26)$$

$$g_{00}^{(AOD)} = -e^{T_0(t)} f_{\text{in}}(r), \quad g_{11}^{(AOD)} = f_{\text{in}}^{-1}(r), \quad (27)$$

$$g_{00}^{(BOD)} = -e^{T_2(t)} f_{21}(r), \quad g_{11}^{(BOD)} = f_{21}^{-1}(r). \quad (28)$$

Here, $f_{\text{in}}$ and $f_{\text{out}}$ are the same as those in (4) and $f_{12}$ and $f_{21}$ are given by similar expressions:

$$f_{12} = 1 - \frac{2km_{12}}{c^2 r}, \quad f_{21} = 1 - \frac{2km_{21}}{c^2 r}. \quad (29)$$

The mass parameters $m_{\text{out}}$, $m_{\text{in}}$, and $m_{12}$ are assumed to be specified by the initial conditions. The mass param-eter $m_{21}$ formed in the region between the shells after their intersection is to be determined from the dynamics of the process and from the joining conditions at point $(r^*, t^*)$.

Let us first write out the equations of motion for the shells before their intersection at $t^*$. To do this, it will suffice to turn to the equations of motion for one shell, for example, in the form (18), (19) and to take into account the fact that the mass parameters in regions *AOD, COA,* and *COB* are $m_{in}$, $m_{12}$, and $m_{out}$, respectively. In this case, it is convenient to use the equations in the forms (19) and (18) for (inner) shell 1 and (outer) shell 2, respectively:

$$\sqrt{f_{12}(r_1) + \left(\frac{dr_1}{cd\tau_1}\right)^2} = \frac{m_{12} - m_{in}}{M_1(r_1)} - \frac{kM_1(r_1)}{2c^2 r_1}, \quad (30)$$

$$\sqrt{f_{12}(r_2) + \left(\frac{dr_2}{cd\tau_2}\right)^2} = \frac{m_{out} - m_{12}}{M_2(r_2)} + \frac{kM_2(r_2)}{2c^2 r_2}, \quad (31)$$

$$M_1 = \sqrt{m_1^2 + \frac{L_1^2}{c^2 r_1^2}}, \quad M_2 = \sqrt{m_2^2 + \frac{L_2^2}{c^2 r_2^2}}. \quad (32)$$

Here, $\tau_1$ and $\tau_2$ are the proper times of the first and second shells and $r_1(\tau_1) = R_1[t(\tau_1)]$ and $r_2(\tau_2) = R_2[t(\tau_2)]$. These equations must be supplemented with the joining conditions for the intervals on both shells. Joining on the first shell (on curve *AO*) yields

$$e^{T_1(t)} f_{12}(r_1)\left(\frac{dt}{d\tau_1}\right)^2 - f_{12}^{-1}(r_1)\left(\frac{dr_1}{cd\tau_1}\right)^2 = 1, \quad (33)$$

$$e^{T_0(t)} f_{in}(r_1)\left(\frac{dt}{d\tau_1}\right)^2 - f_{in}^{-1}(r_1)\left(\frac{dr_1}{cd\tau_1}\right)^2 = 1. \quad (34)$$

Joining on the second shell (on curve *CO*) yields

$$f_{out}(r_2)\left(\frac{dt}{d\tau_2}\right)^2 - f_{out}^{-1}(r_2)\left(\frac{dr_2}{cd\tau_2}\right)^2 = 1, \quad (35)$$

$$e^{T_1(t)} f_{12}(r_2)\left(\frac{dt}{d\tau_2}\right)^2 - f_{12}^{-1}(r_2)\left(\frac{dr_2}{cd\tau_2}\right)^2 = 1. \quad (36)$$

If all free parameters ($m_{in}$, $m_{out}$, $m_{12}$, $m_1$, $m_2$, $L_1$, $L_2$) and initial data to Eqs. (30)–(32) were specified and if the functions $r_1(\tau_1)$ and $r_2(\tau_2)$ were derived, then their substitution into (33)–(36) gives the functions $\tau_1(t)$, $\tau_2(t)$ and $T_1(t)$, $T_0(t)$, i.e., all that is required to determine the motion of the shells before their intersection. The intersection time $t^*$ corresponds to some proper times $\tau_1(t^*)$ and $\tau_2(t^*)$. Therefore, the coordinates of the point of intersection $r^*$ and $t^*$ can be found from two relations,

$$r^* = r_1(\tau_1(t^*)), \quad r^* = r_2(\tau_2(t^*)). \quad (37)$$

Of course, we assume that Eqs. (37) have a solution (the cases where there is no solution correspond to the motion without intersections but we are not interested in such possibilities here).

The equations of motion for the shells after the intersection time $t^*$ can be easily set up in just the same way again by turning to Eqs. (18) and (19) and by taking into account the fact that a new parameter $m_{21}$ emerges in region *BOD*. We use Eq. (18) for (now outer) shell 1 and Eq. (19) for (now inner) shell 2:

$$\sqrt{f_{21}(r_1) + \left(\frac{dr_1}{cd\tau_1}\right)^2} = \frac{m_{out} - m_{21}}{M_1(r_1)} + \frac{kM_1(r_1)}{2c^2 r_1}, \quad (38)$$

$$\sqrt{f_{21}(r_2) + \left(\frac{dr_2}{cd\tau_2}\right)^2} = \frac{m_{21} - m_{in}}{M_2(r_2)} - \frac{kM_2(r_2)}{2c^2 r_2}. \quad (39)$$

Here, $M_1(r_1)$ and $M_2(r_2)$ are given by the same expressions (32) but, naturally, with the values of the functions $r_1(\tau_1)$ and $r_2(\tau_2)$ (and the times $\tau_1$ and $\tau_2$) after the intersection.

Joining the intervals on the first shell (on curve *OB*) yields

$$f_{out}(r_1)\left(\frac{dt}{d\tau_1}\right)^2 - f_{out}^{-1}(r_1)\left(\frac{dr_1}{cd\tau_1}\right)^2 = 1, \quad (40)$$

$$e^{T_2(t)} f_{21}(r_1)\left(\frac{dt}{d\tau_1}\right)^2 - f_{21}^{-1}(r_1)\left(\frac{dr_1}{cd\tau_1}\right)^2 = 1. \quad (41)$$

Joining on the second shell (on curve *OD*) requires that the following relations hold:

$$e^{T_2(t)} f_{21}(r_2)\left(\frac{dt}{d\tau_2}\right)^2 - f_{21}^{-1}(r_2)\left(\frac{dr_2}{cd\tau_2}\right)^2 = 1, \quad (42)$$

$$e^{T_0(t)} f_{in}(r_2)\left(\frac{dt}{d\tau_2}\right)^2 - f_{in}^{-1}(r_2)\left(\frac{dr_2}{cd\tau_2}\right)^2 = 1. \quad (43)$$

We see from Eqs. (38)–(43) that, if the parameter $m_{21}$ were known, then the evolution of the shells after their intersection (for $t > t^*$) would be completely determined by their evolution before the intersection, because the initial data to Eqs. (38) and (39) have already been specified (it is known that the shell positions at time $t^*$ are $r_1 = r_2 = r^*$ and the coordinates of the point of intersection $r^*$ and $t^*$ have already been found from the previous evolution). Thus, we must have an additional physical condition from which we could determine $m_{21}$.

Actually, we have not yet used the continuity condition for the relative velocity of the shells when passing through the point of intersection. We can make sure that this condition is necessary as follows. It is well known that the motion of the shells can be described in coordinates in which all metric coefficients are everywhere continuous and only their first derivatives together with the $\Gamma$ symbols undergo discontinuities (like finite jumps). This implies that the shell accelerations in these coordinates can only have discontinuities like finite jumps and, hence, the velocities must be everywhere continuous. It thus follows that the invariants expressed in terms of the velocities and metric coefficients alone will be everywhere continuous not only in these coordinates but also in any other coordinates, in particular, in our coordinates $r$ and $t$. If there is only one shell, then only one such invariant exists, the norm of the 4-velocity vector for the shell. In the coordinates in which the metric is everywhere continuous, this norm is defined relative to the metric at the points of the shell trajectory themselves and is assumed to be a unit norm. In any other coordinates, the limiting values of the norm as the shell is approached on both of its sides remain unit ones. This circumstance is expressed by relations (6) and (7), which may be called the joining conditions.

In the presence of two shells, the aforesaid remains valid for each of them, which is expressed by the joining conditions (33)–(36) and (40)–(43). However, we now have one more invariant that can be made up only from the velocities and metric coefficients, namely, the scalar product of the unit 4-velocity vectors for the first and second shells. Before the intersection (trajectories $CO$ and $AO$), this quantity can be determined everywhere in the region between the shells (sector $COA$) by making an appropriate parallel transport of the 4-velocity vectors for both shells to the points of this region from the shell sides adjacent to sector $COA$. We emphasize that this scalar product cannot be determined in such a way in sectors $COB$ and $AOD$, because, in this case, we would have to transport the 4-velocity vector for one of the shells through the trajectory of the other shell, i.e., through the point of discontinuities in the $\Gamma$ symbols, and this parallel transport is not uniquely determined. After the intersection, the scalar product of the unit 4-velocity vectors for the shells is uniquely determined for the same reason only in the region between the shells, i.e., in sector $BOD$. Actually, there is no need to make all these parallel transports, because both vectors are at the same point of spacetime at the intersection time, and we are interested only in this point. The previous analysis serves only to show that the limit of the scalar product of the unit 4-velocity vectors for the shells at the point of intersection $O$ should be calculated relative to the metric in sector $COA$ when it is approached from the side "before the intersection" (i.e., at $t = t^* - 0$) and relative to the metric in sector $BOD$ when point $O$ is approached on the side "after the intersection" (i.e., at $t = t^* + 0$). The continuity of this scalar product in the coordinate system where this metric is continuous and its invariance require that these limits also be equal in our coordinates. This is precisely the additional condition from which the parameter $m_{21}$ can be determined. Let us now derive this condition in an explicit form. The unit tangent vector to trajectory $AO$ is

$$u^i_{AO} = (u^0_{AO}, u^1_{AO}, u^2_{AO}, u^3_{AO}) = \left(\frac{dt}{d\tau_1}, \frac{dr_1}{cd\tau_1}, 0, 0\right)_{t \leq t^*}, \quad (44)$$

and the unit tangent vector to trajectory $CO$ is

$$u^i_{CO} = (u^0_{CO}, u^1_{CO}, u^2_{CO}, u^3_{CO}) = \left(\frac{dt}{d\tau_2}, \frac{dr_2}{cd\tau_2}, 0, 0\right)_{t \leq t^*}. \quad (45)$$

The fact that these are actually unit vectors follows from the joining equations (33) and (36).

The components of vector (44) can be easily expressed from Eqs. (30) and (33) as

$$\left(\frac{dt}{d\tau_1}\right)_{t \leq t^*} = \frac{1}{f_{12}(r_1)} e^{-T_1(t)/2} \left[\frac{m_{12} - m_{\text{in}}}{M_1(r_1)} - \frac{kM_1(r_1)}{2c^2 r_1}\right], \quad (46)$$

$$\left(\frac{dr_1}{cd\tau_1}\right)_{t \leq t^*}$$
$$= \delta_1 \sqrt{\left[\frac{m_{12} - m_{\text{in}}}{M_1(r_1)} - \frac{kM_1(r_1)}{2c^2 r_1}\right]^2 - f_{12}(r_1)}, \quad (47)$$

where

$$\delta_1 = \text{sgn}\left(\frac{dr_1}{cd\tau_1}\right)_{t \leq t^*}. \quad (48)$$

For the components of vector (45), we obtain the following expressions from Eqs. (31) and (36):

$$\left(\frac{dt}{d\tau_2}\right)_{t \leq t^*}$$
$$= \frac{1}{f_{12}(r_2)} e^{-T_1(t)/2} \left[\frac{m_{\text{out}} - m_{12}}{M_2(r_2)} + \frac{kM_2(r_2)}{2c^2 r_2}\right], \quad (49)$$

$$\left(\frac{dr_2}{cd\tau_2}\right)_{t \leq t^*}$$
$$= \delta_2 \sqrt{\left[\frac{m_{\text{out}} - m_{12}}{M_2(r_2)} + \frac{kM_2(r_2)}{2c^2 r_2}\right]^2 - f_{12}(r_2)}, \quad (50)$$

$$\delta_2 = \text{sgn}\left(\frac{dr_2}{cd\tau_2}\right)_{t \leq t^*}. \quad (51)$$

Using the metric (26) in region $COA$, we now need to calculate the quantity

$$Q = \{g^{(COA)}_{00} u^0_{AO} u^0_{CO} + g^{(COA)}_{11} u^1_{AO} u^1_{CO}\}_{t = t^*, r = r_1 = r_2 = r^*}. \quad (52)$$

From the preceding results, we obtain

$$Q = \frac{1}{f_{12}(r^*)}\Biggl\{-\Biggl[\frac{m_{12}-m_{\text{in}}}{M_1(r^*)} - \frac{kM_1(r^*)}{2c^2 r^*}\Biggr] \\ \times \Biggl[\frac{m_{\text{out}}-m_{12}}{M_2(r^*)} + \frac{kM_2(r^*)}{2c^2 r^*}\Biggr] \\ + \delta_1\delta_2\sqrt{\Biggl[\frac{m_{12}-m_{\text{in}}}{M_1(r^*)} - \frac{kM_1(r^*)}{2c^2 r^*}\Biggr]^2 - f_{12}(r^*)} \\ \times \sqrt{\Biggl[\frac{m_{\text{out}}-m_{12}}{M_2(r^*)} + \frac{kM_2(r^*)}{2c^2 r^*}\Biggr]^2 - f_{12}(r^*)}\Biggr\}. \quad (53)$$

Let us now turn to the region after the intersection time. For the unit tangent vectors to trajectories $OB$ and $OD$, we have

$$u_{OB}^i = (u_{OB}^0, u_{OB}^1, u_{OB}^2, u_{OB}^3) \\ = \left(\frac{dt}{d\tau_1}, \frac{dr_1}{cd\tau_1}, 0, 0\right)_{t \geq t^*}, \quad (54)$$

$$u_{OD}^i = (u_{OD}^0, u_{OD}^1, u_{OD}^2, u_{OD}^3) \\ = \left(\frac{dt}{d\tau_2}, \frac{dr_2}{cd\tau_2}, 0, 0\right)_{t \geq t^*}; \quad (55)$$

we see from the joining conditions (41) and (42) that these are actually unit vectors. For the components of vector (54), we obtain expressions from Eqs. (38) and (41) similar to (46) and (47) with the substitutions $f_{21}(r_1)$ for $f_{12}(r_1)$, $m_{21}$ for $m_{12}$, $m_{\text{out}}$ for $m_{\text{in}}$, $T_2(t)$ for $T_1(t)$, $\delta_1'$ for $\delta_1$, $[-M_1(r_1)]$ for $M_1(r_1)$, and $t \geq t^*$ for $t \leq t^*$. The components of vector (55) follow from Eqs. (39) and (42); the expressions for them can be derived from (49) and (50) by substituting $f_{21}(r_2)$ for $f_{12}(r_2)$, $m_{21}$ for $m_{12}$, $m_{\text{in}}$ for $m_{\text{out}}$, $T_2(t)$ for $T_1(t)$, $\delta_2'$ for $\delta_2$, $[-M_2(r_2)]$ for $M_2(r_2)$, and $t \geq t^*$ for $t \leq t^*$.

Accordingly, $\delta_1'$ and $\delta_2'$ are defined as in (48) and (51) but for $t \geq t^*$. Using the metric (28) in region $BOD$, let us calculate the scalar product

$$Q' = \{g_{00}^{(BOD)} u_{OB}^0 u_{OD}^0 + g_{11}^{(BOD)} u_{OB}^1 u_{OD}^1\}_{t=t^*, r=r_1=r_2=r^*}. \quad (56)$$

We easily obtain

$$Q' = \frac{1}{f_{21}(r^*)}\Biggl\{-\Biggl[\frac{m_{\text{out}}-m_{21}}{M_1(r^*)} + \frac{kM_1(r^*)}{2c^2 r^*}\Biggr] \\ \times \Biggl[\frac{m_{21}-m_{\text{in}}}{M_2(r^*)} - \frac{kM_2(r^*)}{2c^2 r^*}\Biggr] \\ + \delta_1'\delta_2'\sqrt{\Biggl[\frac{m_{21}-m_{\text{in}}}{M_2(r^*)} - \frac{kM_2(r^*)}{2c^2 r^*}\Biggr]^2 - f_{21}(r^*)} \\ \times \sqrt{\Biggl[\frac{m_{\text{out}}-m_{21}}{M_1(r^*)} + \frac{kM_1(r^*)}{2c^2 r^*}\Biggr]^2 - f_{21}(r^*)}\Biggr\}. \quad (57)$$

The necessary continuity condition is now the requirement that

$$Q = Q', \quad (58)$$

where $Q$ and $Q'$ are given by (53) and (57), respectively. Since the intersection coordinate $r^*$ is assumed to be known, condition (58) is the equation for determining the parameter $m_{21}$.

We began the derivation of Eq. (58) by calling it the continuity condition for the relative velocity of the shells at point $(r^*, t^*)$ in advance. Let us explain this in more detail. We define the "physical" shell velocities $v_1$ and $v_2$ in region $COA$ as

$$\frac{v_1^2}{c^2} = \frac{g_{11}^{(COA)}(r_1) dr_1^2}{-g_{00}^{(COA)}(r_1) c^2 dt^2}, \\ \frac{v_2^2}{c^2} = \frac{g_{11}^{(COA)}(r_2) dr_2^2}{-g_{00}^{(COA)}(r_2) c^2 dt^2}. \quad (59)$$

Using the joining equations (33) and (36), we then obtain

$$g_{11}^{(COA)}(r_1)\left(\frac{dr_1}{cd\tau_1}\right)^2 = \frac{v_1^2/c^2}{1-v_1^2/c^2}, \\ g_{00}^{(COA)}(r_1)\left(\frac{dt}{d\tau_1}\right)^2 = -\frac{1}{1-v_1^2/c^2}, \quad (60)$$

$$g_{11}^{(COA)}(r_2)\left(\frac{dr_2}{cd\tau_2}\right)^2 = \frac{v_2^2/c^2}{1-v_2^2/c^2}, \\ g_{00}^{(COA)}(r_2)\left(\frac{dt}{d\tau_2}\right)^2 = -\frac{1}{1-v_2^2/c^2}. \quad (61)$$

It now follows from (44), (45), and (52) that

$$Q = \left\{\frac{v_1 v_2/c^2 - 1}{\sqrt{1-v_1^2/c^2}\sqrt{1-v_2^2/c^2}}\right\}_{t=t^*, r_1=r_2=r^*}. \quad (62)$$

The "physical" velocities of the shells after their intersection, $v_1'$ and $v_2'$, are defined similarly:

$$\frac{v_1'^2}{c^2} = \frac{g_{11}^{(BOD)}(r_1) dr_1^2}{-g_{00}^{(BOD)}(r_1) c^2 dt^2}, \\ \frac{v_2'^2}{c^2} = \frac{g_{11}^{(BOD)}(r_2) dr_2^2}{-g_{00}^{(BOD)}(r_2) c^2 dt^2}. \quad (63)$$

We then find from the joining equations (41) and (42) that

$$g_{11}^{BOD}(r_1)\left(\frac{dr_1}{cd\tau_1}\right)^2 = \frac{v_1'^2/c^2}{1 - v_1'^2/c^2},$$

$$g_{00}^{BOD}(r_1)\left(\frac{dt}{d\tau_1}\right)^2 = -\frac{1}{1 - v_1'^2/c^2}, \quad (64)$$

$$g_{11}^{BOD}(r_2)\left(\frac{dr_2}{cd\tau_2}\right)^2 = \frac{v_2'^2/c^2}{1 - v_2'^2/c^2},$$

$$g_{00}^{BOD}(r_2)\left(\frac{dt}{d\tau_2}\right)^2 = -\frac{1}{1 - v_2'^2/c^2}. \quad (65)$$

It follows from (54)–(56) that

$$Q' = \left\{\frac{v_1' v_2'/c^2 - 1}{\sqrt{1 - v_1'^2/c^2}\sqrt{1 - v_2'^2/c^2}}\right\}_{t=t^*, r_1=r_2=r^*}. \quad (66)$$

Naturally, the continuity of $Q$ when passing through the point of intersection ($r^*, t^*$) also implies the continuity of any function of $Q$, in particular, of

$$\frac{\sqrt{Q^2 - 1}}{Q} = \frac{v_1/c - v_2/c}{1 - v_1 v_2/c^2}. \quad (67)$$

The latter is nothing else but the relative velocity of two "particles," as it is defined in relativistic mechanics. This gives grounds to call condition (58) the continuity condition for the relative velocity of the shells.

To work with Eq. (58), it is convenient first to simplify the form of expressions (53) and (57) for $Q$ and $Q'$ by denoting

$$\sigma_1 = \left[\frac{kM_1(r^*)}{c^2 r^*}\right]^2, \quad \sigma_2 = \left[\frac{kM_2(r^*)}{c^2 r^*}\right]^2 \quad (68)$$

and by expressing the differences between the mass parameters in $Q$ and $Q'$ using formulas (4) and (29) in terms of the differences between the functions $f(r)$ taken at point $r = r^*$ in accordance with the relation

$$m_a - m_b = \frac{c^2 r^*}{2k}[f_b(r^*) - f_a(r^*)], \quad (69)$$

where $a$ and $b$ mean the indices "in," "out," 12, and 21. Now, $Q$ and $Q'$ can be represented as

$$Q = -(4\sqrt{\sigma_1 \sigma_2} f_{12})^{-1}$$
$$\times [(f_{in} - f_{12} - \sigma_1)(f_{12} - f_{out} + \sigma_2)$$
$$- \delta_1 \delta_2 \sqrt{(f_{in} - f_{12} - \sigma_1)^2 - 4\sigma_1 f_{12}}$$
$$\times \sqrt{(f_{12} - f_{out} + \sigma_2)^2 - 4\sigma_2 f_{12}}], \quad (70)$$

$$Q' = -(4\sqrt{\sigma_1 \sigma_2} f_{21})^{-1}$$
$$\times [(f_{21} - f_{out} + \sigma_1)(f_{in} - f_{21} - \sigma_2)$$
$$- \delta_1' \delta_2' \sqrt{(f_{21} - f_{out} + \sigma_1)^2 - 4\sigma_1 f_{21}}$$
$$\times \sqrt{(f_{in} - f_{21} - \sigma_2)^2 - 4\sigma_2 f_{21}}], \quad (71)$$

where we omitted the argument $r^*$ in the functions $f_a$ to save space. In all the subsequent formulas given below, the quantities $f_a$ without any argument will mean their values at $r = r^*$.

If the equation $Q = Q'$ is written using expressions (70) and (71) and if the first term from $Q'$ is carried over to the left-hand side of this equation, then squaring the resulting relation yields a quadratic equation for the unknown $f_{21}$, i.e., for the parameter $m_{21}$. This procedure is somewhat cumbersome, and, in addition, it does not answer the question of which of the two solutions should be chosen. However, there is a method not only to easily find a solution to the equation $Q = Q'$ in a simple form but also to find this solution unambiguously and directly from the continuity equations without resorting to any additional physical considerations. The point is that, apart from Eq. (58), there are other similar continuity requirements when passing through the point of intersection of the shells; their use makes it easier to find the solution and singles it out unequivocally. As we saw, Eq. (58) was derived by requiring that the scalar product of the unit 4-velocity vectors for the shells be continuous when passing through the point of their intersection from sector $COA$ (before intersection) into sector $BOD$ (after intersection). It is easy to see, however, that all the reasoning that accompanied the derivation of Eq. (58) is also equally applicable to the passage through the point of intersection from sector $AOD$ into sector $COB$; i.e., we may also assert that the limit of the scalar product of the unit tangent vectors to trajectories $AO$ and $OD$ when point $O$ is approached from sector $AOD$ must be equal to the limit of the scalar product of the unit tangent vectors to trajectories $OB$ and $CO$ when point $O$ is approached from sector $COB$. This continuity condition can be easily obtained in an explicit form by repeating a procedure similar to the procedure that led to Eq. (58). We first calculate the scalar product

$$P = \{g_{00}^{(AOD)} u_{AO}^0 u_{OD}^0 + g_{11}^{(AOD)} u_{AO}^1 u_{OD}^1\}_{t=t^*, r=r_1=r_2=r^*} \quad (72)$$

using the equations of motion for shells $AO$ and $OD$ in the form (18) and the joining conditions (34) and (43) and then calculate the scalar product

$$P' = \{g_{00}^{(COB)} u_{CO}^0 u_{OB}^0 + g_{11}^{(COB)} u_{CO}^1 u_{OB}^1\}_{t=t^*, r=r_1=r_2=r^*} \quad (73)$$

using the equations of motion for shells *CO* and *OB* in the form (19) and the joining conditions (35) and (40). Subsequently, we equate the results:

$$P' = P. \tag{74}$$

In the same notation that we used to derive expressions (70) and (71), the quantities $P$ and $P'$ are

$$P = -(4\sqrt{\sigma_1\sigma_2}f_{\text{in}})^{-1}$$
$$\times [(f_{\text{in}} - f_{12} + \sigma_1)(f_{\text{in}} - f_{21} + \sigma_2)$$
$$- \delta_1\delta_2'\sqrt{(f_{\text{in}} - f_{12} + \sigma_1)^2 - 4\sigma_1 f_{\text{in}}} \tag{75}$$
$$\times \sqrt{(f_{\text{in}} - f_{21} + \sigma_2)^2 - 4\sigma_2 f_{\text{in}}}],$$

$$P' = -(4\sqrt{\sigma_1\sigma_2}f_{\text{out}})^{-1}$$
$$\times [(f_{21} - f_{\text{out}} - \sigma_1)(f_{12} - f_{\text{out}} - \sigma_2)$$
$$- \delta_1'\delta_2\sqrt{(f_{21} - f_{\text{out}} - \sigma_1)^2 - 4\sigma_1 f_{\text{out}}} \tag{76}$$
$$\times \sqrt{(f_{12} - f_{\text{out}} - \sigma_2)^2 - 4\sigma_2 f_{\text{out}}}].$$

In addition, it is clear that all that we have said above about the continuity conditions at the point of intersection is not restricted only to the passages from *COA* into *BOD* and from *AOD* into *COB* but is equally applicable to the passages in general from any sector into any other sector. This implies that all four quantities $Q$, $Q'$, $P$, and $P'$ must actually be identical and, in addition to conditions (58) and (74), we must require that one more equation, for example, $P = Q$, be satisfied. Thus, the complete set of continuity conditions at the point of intersection can be written as

$$Q = Q', \quad Q = P, \quad Q = P'. \tag{77}$$

It turns out that these three quadratic equations for one unknown $f_{21}$ actually have one single common root. This root can be easily determined[3] from the first two equations from (77) and is

$$f_{21} = f_{\text{out}} + f_{\text{in}} - f_{12} - (2f_{12})^{-1}$$
$$\times [(f_{\text{in}} - f_{12} - \sigma_1)(f_{12} - f_{\text{out}} + \sigma_2)$$
$$- \delta_1\delta_2\sqrt{(f_{\text{in}} - f_{12} - \sigma_1)^2 - 4\sigma_1 f_{12}} \tag{78}$$
$$\times \sqrt{(f_{12} - f_{\text{out}} + \sigma_2)^2 - 4\sigma_2 f_{12}}].$$

---

[3] This requires expressing the components of the unit 4-velocity vectors in terms of hyperbolic functions. Thus, for example, if we introduce the angles $\alpha_1$ and $\alpha_2$ according to $\delta_1\sqrt{(f_{\text{in}} - f_{12} - \sigma_1)^2 - 4\sigma_1 f_{12}} = \sqrt{4\sigma_1 f_{12}}\sinh\alpha_1$ and $\delta_2\sqrt{(f_{12} - f_{\text{out}} + \sigma_2)^2 - 4\sigma_2 f_{12}} = \sqrt{4\sigma_2 f_{12}}\sinh\alpha_2$, then $Q = -\cosh(\alpha_1 - \alpha_2)$. Similar notation should be introduced for the quantities from which the scalar products $P$, $P'$, and $Q'$ are made up. Equation (77) can then be easily resolved for the "angles" that contain the unknown $f_{21}$.

The third equation from (77) after substituting the expression for $f_{21}$ in it is satisfied identically.

Formula (78) solves the problem of determining the mass parameter $m_{21}$ from the quantities specified at the evolutionary stage before intersection. The energy transfer between the shells as they intersect is determined along with it. Indeed, the total conserved energies of the shells before their intersection are defined as

$$E_1 = (m_{12} - m_{\text{in}})c^2, \quad E_2 = (m_{\text{out}} - m_{12})c^2, \tag{79}$$

where $E_1$ is the energy of (inner) shell 1 and $E_2$ is the energy of (outer) shell 2. During the intersection, the energy of each shell undergoes a discontinuity; subsequently, their energies become equal to $E_1'$ and $E_2'$, respectively:

$$E_1' = (m_{\text{out}} - m_{21})c^2, \quad E_2' = (m_{21} - m_{\text{in}})c^2. \tag{80}$$

The conservation of total energy of the shells when passing through the point of intersection automatically follows from these expressions:

$$E_1 + E_2 = E_1' + E_2'. \tag{81}$$

Using this relation, we can write the energies of the shells after their intersection as

$$E_1' = E_1 - \Delta E, \quad E_2' = E_2 + \Delta E. \tag{82}$$

It follows from (80), (81), and (69) that

$$\Delta E = \frac{c^4 r^*}{2k}(f_{\text{in}} + f_{\text{out}} - f_{12} - f_{21}). \tag{83}$$

Substituting the solution (78) for $f_{21}$ in this expression yields

$$\Delta E = \frac{c^4 r^*}{4k f_{12}}[(f_{\text{in}} - f_{12} - \sigma_1)(f_{12} - f_{\text{out}} + \sigma_2)$$
$$- \delta_1\delta_2\sqrt{(f_{\text{in}} - f_{12} - \sigma_1)^2 - 4\sigma_1 f_{12}} \tag{84}$$
$$\times \sqrt{(f_{12} - f_{\text{out}} + \sigma_2)^2 - 4\sigma_2 f_{12}}].$$

Note that the square bracket multiplied by the factor $f_{12}^{-1}$ in formulas (78) and (84) can be easily expressed in terms of the scalar product $Q$ [see formula (70)], which, in turn, can be easily expressed in terms of the shell velocities $v_1$ and $v_2$ at the point of intersection [relation (80)]. This representation of $\Delta E$ is

$$\Delta E = -\frac{kM_1(r^*)M_2(r^*)}{r^*}Q$$
$$= \left(\frac{kM_1 M_2}{r} \frac{1 - v_1 v_2/c^2}{\sqrt{1 - v_1^2/c^2}\sqrt{1 - v_2^2/c^2}}\right)_{r = r^*}, \tag{85}$$

and it is convenient in some cases (in particular, for obtaining the nonrelativistic approximation). The cor-responding expression for the metric coefficient $f_{21}$ can be written as

$$f_{21} = f_{\text{out}} + f_{\text{in}} - f_{12} - \frac{2k}{c^4 r^*}\Delta E. \qquad (86)$$

## 4. THE NONRELATIVISTIC (NEWTONIAN) APPROXIMATION

In the nonrelativistic approximation, the total energies of the shells (79) and (80) can be expanded as

$$E = mc^2 + \mathcal{E} + o(1/c^2),$$

where $m$ and $\mathcal{E}$ do not depend on $c$. This means that the differences between the mass parameters, to within $1/c^2$ inclusive, are

$$m_{12} - m_{\text{in}} = m_1 + \frac{\mathcal{E}_1}{c^2}, \quad m_{\text{out}} - m_{12} = m_2 + \frac{\mathcal{E}_2}{c^2}, \qquad (87)$$

$$m_{\text{out}} - m_{21} = m_1 + \frac{\mathcal{E}'_1}{c^2}, \quad m_{21} - m_{\text{in}} = m_2 + \frac{\mathcal{E}'_2}{c^2}. \qquad (88)$$

Here, $m_1$ and $m_2$ are the rest masses of the shells, those that appear in formulas (32) for the effective masses $M_1$ and $M_2$. The quantities $\mathcal{E}$ in (87) and (88) are the total nonrelativistic energies of the shells; the law of their conservation follows from (87) and (88):

$$\mathcal{E}_1 + \mathcal{E}_2 = \mathcal{E}'_1 + \mathcal{E}'_2. \qquad (89)$$

Relation (82) now takes the form

$$\mathcal{E}'_1 = \mathcal{E}_1 - \Delta\mathcal{E}, \quad \mathcal{E}'_2 = \mathcal{E}_2 + \Delta\mathcal{E}, \qquad (90)$$

where $\Delta\mathcal{E} = (\Delta E)_{c=\infty}$ follows from expression (85). Since $M_1 = m_1$ and $M_2 = m_2$ in the first nonvanishing order in $1/c^2$, we obtain the following nonrelativistic formulas for energy transfer during the intersection from (85):

$$\Delta\mathcal{E} = \frac{km_1 m_2}{r^*}. \qquad (91)$$

The nonrelativistic equations of motion for the shells before their intersection can be easily derived from the exact equations (30)–(32). The proper times $\tau_1$ and $\tau_2$ in the principal order are equal to the global time $t$, so $(dr_1/cd\tau_1)^2$ and $(dr_2/cd\tau_2)^2$ in these equations are nothing else but $v_1^2/c^2$ and $v_2^2/c^2$, respectively, where

$$v_1 = \frac{dr_1}{dt}, \quad v_2 = \frac{dr_2}{dt}. \qquad (92)$$

Expanding Eqs. (30)–(32) up to the order $1/c^2$ inclusive yields

$$\mathcal{E}_1 = \frac{m_1 v_1^2}{2} - \frac{km_1(m_{\text{in}} + m_1/2)}{r_1} + \frac{L_1^2}{2r_1^2 m_1}, \qquad (93)$$

$$\mathcal{E}_2 = \frac{m_2 v_2^2}{2} - \frac{km_2(m_{\text{in}} + m_2/2 + m_1)}{r_2} + \frac{L_2^2}{2r_2^2 m_2}. \qquad (94)$$

A similar operation with Eqs. (38) and (39) leads to the following equations of motion for the shells after their intersection:

$$\mathcal{E}'_1 = \frac{m_1 v_1^2}{2} - \frac{km_1(m_{\text{in}} + m_1/2 + m_2)}{r_1} + \frac{L_1^2}{2r_1^2 m_1}, \qquad (95)$$

$$\mathcal{E}'_2 = \frac{m_2 v_2^2}{2} - \frac{km_2(m_{\text{in}} + m_2/2)}{r_2} + \frac{L_2^2}{2r_2^2 m_2}. \qquad (96)$$

Together with relations (90) and (91) and the initial data $r_1 = r_2 = r^*$ at $t = t^*$, Eqs. (95) and (96) completely determine the evolution of the shells immediately after $t^*$ (i.e., before the next possible intersection).

It follows from Eqs. (93)–(96) applied to the point $r_1 = r_2 = r^*$ that

$$\mathcal{E}_1 - \mathcal{E}'_1 = \frac{m_1}{2}[v_1^2(r^*) - v_1'^2(r^*)] + \frac{km_1 m_2}{r^*},$$

$$\mathcal{E}_2 - \mathcal{E}'_2 = \frac{m_2}{2}[v_2^2(r^*) - v_2'^2(r^*)] - \frac{km_1 m_2}{r^*}. \qquad (97)$$

Here, we denoted the limits of the velocities $v_{1,2}$ at point $r^*$ from the side "after intersection" by $v'_{1,2}(r^*)$, while $v_{1,2}(r^*)$ are the limits of these velocities at point $r^*$ from the side "before intersection." Relations (97) and the law of change in energy (90) and (91) show that the shell velocities are continuous at the point of intersection in the nonrelativistic approximation:

$$v'_1(r^*) = v_1(r^*), \quad v'_2(r^*) = v_2(r^*) \qquad (98)$$

(the cases where the velocities change sign are excluded from the additional physical requirements for joining the evolutions discussed in the Introduction). Of course, condition (98) is obvious in advance and requires no discussion. We deduced it here only to show the consistency of the entire procedure. Note that, when studying the Newtonian approximation, we can begin directly from condition (98) as the main postulate (as we did in [8]).

The exact formula (84) for $\Delta E$ written in terms of the velocities in the form (85) also allows the post-Newtonian correction to the energy transfer to be easily calculated. Expanding the right-hand side of Eq. (85) to within order $1/c^2$ inclusive yields

$$\Delta E = \frac{km_1m_2}{r^*} + \frac{1}{2c^2}\left\{\frac{km_1m_2}{r^*}[v_1(r^*) - v_2(r^*)]^2 \right. \quad (99)$$
$$\left. + \frac{km_2L_1^2}{m_1r^{*3}} + \frac{km_1L_2^2}{m_2r^{*3}}\right\}.$$

The velocities $v_1(r^*)$ and $v_2(r^*)$ in this formula should be determined from Eqs. (93) and (94) applied to the point $r_1 = r_2 = r^*$; i.e., these are the standard velocities of the Newtonian approximation. Of course, if we take into consideration the post-Newtonian correction to $\Delta E$, then we must take into account such corrections to all energies $E$; i.e., we must write out the next terms of the expansion in $1/c^2$ in energies after the Newtonian ones (93)–(96). We performed the corresponding calculations but do not present their results here, because they are relatively cumbersome and are not needed at the current stage of our studies.

## 5. SHELLS WITH ZERO EFFECTIVE REST MASSES

If the shell particles move only radially and have a zero rest mass, then $m_1 = m_2 = 0$ and $L_1 = L_2 = 0$. In this case, both shells have a zero effective rest mass, i.e., $M_1 = M_2 = 0$. As previously, the shell intersection corresponds to Fig. 1, with the only difference that trajectories $CO$, $AO$, $OB$, and $OD$ are now isotropic. In all four sectors, the metric is given by the same formulas (25)–(28), in which the functions $f$ have the same form (4), (29). As previously, the problem consists in determining the radius $r^*$ of the point of intersection of the shells and then the mass parameter $m_{21}$ or, equivalently, $f_{21}(r^*)$. Naturally, the latter must follow from formula (78), in which we should set $\sigma_1 = \sigma_2 = 0$ and assume that $\delta_1\delta_2 = -1$, because such $s$-wave lightlike shells can intersect only if they initially move toward each other. Given also that the mass parameters in the physical region are arranged in the order

$$m_{\text{out}} > m_{12} > m_{\text{in}}, \quad m_{\text{out}} > m_{21} > m_{\text{in}}, \quad (100)$$

or, equivalently,

$$f_{\text{out}} < f_{12} < f_{\text{in}}, \quad f_{\text{out}} < f_{21} < f_{\text{in}}, \quad (101)$$

we obtain the following result from (78):

$$f_{21}(r^*)f_{12}(r^*) = f_{\text{in}}(r^*)f_{\text{out}}(r^*). \quad (102)$$

This is nothing else but the relation derived by Dray and 't Hooft [17]; these authors considered the intersection of two light spherically symmetric shells with a purely radial motion of their constituent "photons."

To completely describe the behavior of light shells, we must also set up the equations of their motion. Now,

the proper time $\tau$ cannot be used in these equations, because it does not exist. If there is only one shell, then expressions (1) and (2) for the intervals remain the same, while expression (3) changes to

$$-(ds^2)_{\text{on}} = R_0^2(t)d\Omega^2. \quad (103)$$

Thus, the joining conditions now imply that the radial-time part of the interval must become zero on both sides of the shells; i.e., instead of (6) and (7), we obtain

$$f_{\text{in}}(R_0)e^{T(t)} - f_{\text{in}}^{-1}(R_0)\left(\frac{dR_0}{cdt}\right)^2 = 0, \quad (104)$$

$$f_{\text{out}}(R_0) - f_{\text{out}}^{-1}(R_0)\left(\frac{dR_0}{cdt}\right)^2 = 0. \quad (105)$$

The equation of motion for the shell $r = R_0(t)$ follows from (105),

$$\left(\frac{dR_0}{cdt}\right)^2 = f_{\text{out}}^2(R_0), \quad (106)$$

and Eq. (104) then gives the dilaton factor $e^{T(t)}$:

$$e^{T(t)} = \frac{f_{\text{out}}^2(R_0)}{f_{\text{in}}^2(R_0)}. \quad (107)$$

It is easy to see that the equation of motion in the forms (18)–(20) is now no longer required, because it simply copies Eq. (106). This can be easily shown first (at $\mu \neq 0$) by changing to the global time $t$ instead of $\tau$ in the equations and then passing to the limit $\mu \longrightarrow 0$.

In the case of two shells, the equations of motion for the second (outer) shell before their intersection follow the equality of the radial-time interval to zero on both of its sides:

$$f_{\text{out}}(R_2) - f_{\text{out}}^{-1}(R_2)\left(\frac{dR_2}{cdt}\right)^2 = 0, \quad (108)$$

$$e^{T_1(t)}f_{12}(R_2) - f_{12}^{-1}(R_2)\left(\frac{dR_2}{cdt}\right)^2 = 0, \quad (109)$$

whence it follows that

$$\left(\frac{dR_2}{cdt}\right)^2 = f_{\text{out}}^2(R_2), \quad e^{T_1(t)} = \frac{f_{\text{out}}^2(R_2)}{f_{12}^2(R_2)}. \quad (110)$$

For (inner) shell 1 before the intersection,

$$e^{T_1(t)}f_{12}(R_1) - f_{12}^{-1}(R_1)\left(\frac{dR_1}{cdt}\right)^2 = 0, \quad (111)$$

$$e^{T_0(t)}f_{\text{in}}(R_1) - f_{\text{in}}^{-1}(R_1)\left(\frac{dR_1}{cdt}\right)^2 = 0. \quad (112)$$

Substituting $e^{T_1}$ from (110) in (111) yields

$$\left(\frac{dR_1}{cdt}\right)^2 = \frac{f_{\text{out}}^2(R_2)f_{12}^2(R_1)}{f_{12}^2(R_2)},$$

$$e^{T_0(t)} = \frac{f_{\text{out}}^2(R_2)f_{12}^2(R_1)}{f_{12}^2(R_2)f_{\text{in}}^2(R_1)}. \tag{113}$$

After the intersection, we have for (now outer) shell 1

$$f_{\text{out}}(R_1) - f_{\text{out}}^{-1}(R_1)\left(\frac{dR_1}{cdt}\right)^2 = 0, \tag{114}$$

$$e^{T_2(t)}f_{21}(R_1) - f_{21}^{-1}(R_1)\left(\frac{dR_1}{cdt}\right)^2 = 0, \tag{115}$$

whence it follows that

$$\left(\frac{dR_1}{cdt}\right)^2 = f_{\text{out}}^2(R_1), \quad e^{T_2(t)} = \frac{f_{\text{out}}^2(R_1)}{f_{21}^2(R_1)}. \tag{116}$$

For (inner) shell 2 after the intersection,

$$e^{T_2(t)}f_{21}(R_2) - f_{21}^{-1}(R_2)\left(\frac{dR_2}{cdt}\right)^2 = 0, \tag{117}$$

$$e^{T_0(t)}f_{\text{in}}(R_2) - f_{\text{in}}^{-1}(R_2)\left(\frac{dR_2}{cdt}\right)^2 = 0. \tag{118}$$

Substituting $e^{T_2}$ from (116) into (117) yields

$$\left(\frac{dR_2}{cdt}\right)^2 = \frac{f_{\text{out}}^2(R_1)f_{21}^2(R_2)}{f_{21}^2(R_1)},$$

$$e^{T_0(t)} = \frac{f_{\text{out}}^2(R_1)f_{21}^2(R_2)}{f_{21}^2(R_1)f_{\text{in}}^2(R_2)}. \tag{119}$$

If the initial data to the first of Eqs. (110) and to the first of Eqs. (113) are specified and if the mass parameters $m_{12}$, $m_{\text{in}}$, and $m_{\text{out}}$ are also specified, then the trajectories $r = R_1(t)$ and $r = R_2(t)$ of the shells before their intersection are completely determined together with the point of their intersection $r^* = R_1(t^*) = R_2(t^*)$. The mass parameter $m_{21}$ is then derived from (102), and the trajectories of the shells after their intersection are completely determined from the first Eqs. (116) and (119).

The energy transfer can be easily calculated from (83) by substituting $f_{21}(r^*)$ expressed from relation (102):

$$\Delta E = \frac{2k(m_{12} - m_{\text{in}})(m_{\text{out}} - m_{12})}{r^* f_{12}(r^*)}. \tag{120}$$

Note also that, if we took $\delta_1\delta_2 = 1$, then we would obtain

$$f_{21}(r^*) = f_{\text{in}}(r^*) + f_{\text{out}}(r^*) - f_{12}(r^*), \tag{121}$$

and the energy transfer $\Delta E$ would become zero, reflecting the fact that such light shells moving in the same direction cannot intersect. Indeed, a simple examination of Eqs. (110) and (113) shows that no intersection is possible for $\delta_1\delta_2 = 1$, which, of course, is obvious in advance. At the same time, the existence of solution (121) and, hence, region 21 after the intersection stems in this case from the fact that the exact solution for zero effective masses may be treated as the first approximation for massive shells but in the ultrarelativistic regime, when the effective rest masses $M_1$ and $M_2$ are insignificant and the solution can be expanded in small parameters $\sigma_1$ and $\sigma_2$. In this case, the intersection is possible even if the shells move in the same direction and the energy transfer in the first nonvanishing order is a small quantity linear in $\sigma_1$ and $\sigma_2$, i.e., in $M_1^2$ and $M_2^2$. Our calculation yields

$$\Delta E = \frac{kM_1^2(m_{\text{out}} - m_{12})^2 + kM_2^2(m_{12} - m_{\text{in}})^2}{2r^*(m_{12} - m_{\text{in}})(m_{\text{out}} - m_{12})}, \tag{122}$$

$$\delta_1\delta_2 = 1;$$

the solution (121) for $f_{21}(r^*)$ refers precisely to this situation.

In conclusion, note that, if the photons that constitute the shells move nonradially, then the parameters $L_1$ and $L_2$ are nonzero at zero $m_1$ and $m_2$. In this case, the effective rest masses $M_1$ and $M_2$ are nonzero and the qualitative behavior of such light shells is the same as that of massive shells.

## 6. THE INTERSECTION OF A TEST SHELL WITH A GRAVITATING SHELL

Clearly, the motion and intersections of two test shells are of no interest. Each of them moves as if no other shell exists at all. A nontrivial situation arises when only one of the shells is a test one, while the gravitational field of the other shell is completely taken into account. Naturally, the solution of this problem must follow from the general case using the corresponding passage to the limit. Let us first consider how the passage to the limit of a test shell is accomplished when there is only one shell. To obtain this limit, we must redesignate the constant parameters as

$$m = \lambda m_p, \quad L = \lambda L_p, \quad m_{\text{out}} = m_{\text{in}} + \lambda \frac{E_p}{c^2}, \tag{123}$$

and assume the constants $m_p$, $L_p$, and $E_p$ to be $\lambda$-independent. Subsequently, we must pass to the limit $\lambda \longrightarrow 0$.

In the limit $\lambda \longrightarrow 0$, we have $m_{\text{out}} = m_{\text{in}}$ and $e^{T(t)} = 1$ follows from the joining equations (6) and (7). We now see from (1) and (2) that the metric is the same both inside and outside the shell (as should be the case if it is a test shell):

$$-ds^2 = f_{in}(r)c^2 dt^2 + f_{in}^{-1}(r)dr^2 + r^2 d\Omega^2. \quad (124)$$

Only one relation now remains from the joining conditions:

$$f_{in}(r_0)\left(\frac{dt}{d\tau}\right)^2 - f_{in}^{-1}(r_0)\left(\frac{dr_0}{cd\tau}\right)^2 = 1, \quad (125)$$

which no longer makes sense to call the joining condition. This is simply the condition for normalizing the 4-velocity of a test particle to unity.

We write the effective mass $\mu(\tau)$ as

$$\mu(\tau) = \lambda \mu_p(\tau), \quad (126)$$

and relation (24) then defines $\mu_p$:

$$\mu_p(\tau) = \sqrt{m_p^2 + \frac{L_p^2}{c^2 r_0^2(\tau)}}. \quad (127)$$

Substituting (126) and (127) into Eq. (18) and passing to the limit $\lambda \longrightarrow 0$ yields the following equation of motion for the test shell in the field of the central mass $m_{in}$:

$$\frac{E_p}{c^2} = \mu_p(\tau)\sqrt{f_{in}(r_0) + \left(\frac{dr_0}{cd\tau}\right)^2}. \quad (128)$$

As we see, the zero parameters disappeared from the final equations (124), (125) and (127), (128), and only the finite constants $m_p$, $L_p$, $m_{in}$, and $E_p$ remained; the latter characterize the test shell.

It would be natural to expect that the equation of motion for the test shell must match the equation of a geodesic in the field of a central mass. It turns out that this is actually the case. Using relation (125), we can easily write the equation of motion (128) in the Schwarzschild time $t$ rather than in the proper time:

$$\frac{E_p}{c^2} = \sqrt{m_p^2 + \frac{L_p^2}{c^2 R_0^2(t)}} \sqrt{\frac{f_{in}^3(R_0)}{f_{in}^2(R_0) - \left(\frac{dR_0}{cdt}\right)^2}}. \quad (129)$$

This expression is now easy to compare with that following from the solution of the geodesic equations in the metric (124). For the test shell, Eq. (129) is the only integral of motion in the sense that there are no motions except radial one in the shell as a whole. Therefore, we deal with the match between (129) and the first integral of the geodesic equations that describes the radial part of the motion of the test particle (although the particle, of course, also has a tangential motion). This integral of the geodesic equations is known to exist. It is easy to show that it exactly matches (129) if we identify the total conserved particle energy with $E_p$, the particle rest mass with $m_p$, and the square of the norm of the conserved particle angular momentum with $L_p^2$.

Let us now consider the situation with two shells shown in Fig. 1 and assume that shell 2 (trajectories $CO$ and $OD$) is a test shell. We leave the parameters of shell 1 unchanged and redesignate the parameters of shell 2 in accordance with (123):

$$m_2 = \lambda m_p, \quad L_2 = \lambda L_p, \quad E_2 = \lambda E_p, \quad E_2' = \lambda E_p'.$$

It thus follows that

$$M_2(r) = \lambda M_p(r),$$

where

$$M_p(r) = \sqrt{m_p^2 + \frac{L_p^2}{c^2 r^2}}. \quad (130)$$

Relations (79) and (80) now yield

$$m_{12} = m_{out} - \lambda \frac{E_p}{c^2}, \quad m_{21} = m_{in} + \lambda \frac{E_p'}{c^2},$$

whence we see that the mass parameters on both sides of shell 2 before and after the intersection are equal in the limit $\lambda = 0$:

$$m_{12} = m_{out}, \quad m_{21} = m_{in}. \quad (131)$$

It follows from the joining conditions (35), (36) and (42), (43) in the limit $\lambda = 0$ that

$$e^{T_1} = 1, \quad e^{T_2} = e^{T_0}. \quad (132)$$

Thus, as we see from (25), (26) and (27), (28), the metrics in regions $COB$ and $COA$ are identical, as are the metrics in regions $AOD$ and $BOD$. In other words, the metrics are continuous when passing through trajectories $CO$ and $OD$, as should be the case for a test shell.

Substituting the redesignated parameters in Eqs. (30) and (31) and passing to the limit $\lambda = 0$ yields the following equations of motion for the shells before their intersection:

$$\sqrt{f_{out}(r_1) + \left(\frac{dr_1}{cd\tau_1}\right)^2} = \frac{m_{out} - m_{in}}{M_1(r_1)} - \frac{kM_1(r_1)}{2c^2 r_1}, \quad (133)$$

$$\sqrt{f_{out}(r_2) + \left(\frac{dr_2}{cd\tau_2}\right)^2} = \frac{E_p}{c^2 M_p(r_2)}. \quad (134)$$

Similarly, we derive the equations of motion after their intersection from (38) and (39):

$$\sqrt{f_{in}(r_1) + \left(\frac{dr_1}{cd\tau_1}\right)^2} = \frac{m_{out} - m_{in}}{M_1(r_1)} + \frac{kM_1(r_1)}{2c^2 r_1}, \quad (135)$$

$$\sqrt{f_{in}(r_2) + \left(\frac{dr_2}{cd\tau_2}\right)^2} = \frac{E_p'}{c^2 M_p(r_2)}. \quad (136)$$

The equations of motion (133) and (135) for the gravitating shell 1 are actually the same equation, only the former is written in the form (19), while the latter is written in the form (18); i.e., the entire evolution of shell 1 can be described only by one of these equations extended to both stages of motion, before and after the intersection. Thus, the gravitating shell moves without being affected by the test shell, as should be the case. The situation with the test shell 2 is different. The equations of its motion (134) and (136) are distinctly different. In addition, we must also define its energy $E_p'$ after the intersection via the parameters specified before the intersection. The latter is accomplished by using Eq. (78), in which we must substitute the redesignated parameters and then pass to the limit $\lambda = 0$. This operation leads to the following:

$$E_p' = E_p + \frac{1}{2f_{\text{out}}}[(f_{\text{in}} - f_{\text{out}} - \sigma_1)E_p \\ - \delta_1\delta_2\sqrt{(f_{\text{in}} - f_{\text{out}} - \sigma_1)^2 - 4\sigma_1 f_{\text{out}}} \\ \times \sqrt{E_p^2 - f_{\text{out}}M_p^2 c^4}], \quad (137)$$

where $f_{\text{in}}$, $f_{\text{out}}$, and $M_p$ are also the values of these functions at the point of intersection $r = r^*$.

Equation (137) gives a jump in the energy of the test shell. If necessary, we can also determine the jump in its velocity. To avoid misunderstandings, we first note the following. For the gravitating shell 1, because the equations of its motion before and after the intersection are identical, the derivative $dr_1/d\tau_1$ is continuous at the point of intersection. In contrast, the velocity of this shell defined by formulas (60) and (64) is discontinuous, because

$$g_{11}^{COA}(r^*) \neq g_{11}^{BOD}(r^*).$$

Of course, this discontinuity is not physical and results only from different definitions of the velocity before and after the intersection: the velocity of the gravitating shell is defined with respect to the metric outside it (sector COB) before the intersection and with respect to the metric inside it (sector BOD) after the intersection. It is easy to verify that, if we continued to define the velocity of shell 1 after the intersection with respect to the metric outside it (i.e., in sector COB), then this velocity would be continuous at the point of intersection. The velocity of shell 1 defined everywhere with respect to the metric inside it would also be continuous. Since the gravitating shell does not feel the presence of the test shell, its intersection with the latter is not distinguished in any way and the change in the definition of any quantities at an undistinguished time of evolution, naturally, bears no relation to the physics of the process. Of course, there may be reasons why this strange definition of the velocities is, nevertheless, convenient, but this is another thing altogether. In our study, this was required only to relate the joining conditions at the point of intersection to the continuity of the relative velocity, which may not have been done. Calculating the scalar products $Q$, $Q'$, $P$, and $P'$ and their continuity conditions by no means require introducing any velocities.

The situation with the test shell with the same metric on both of its sides is different. The "physical" velocity of this shell with respect to this metric is unambiguously defined everywhere and cannot have fictitious discontinuities of the type described above. Therefore, the discontinuity in this velocity at the point of intersection is actually connected with physics of the process. Before the intersection, the velocity of the test shell is defined by formulas (61) and (26), in which we should also pass to the limit $\lambda = 0$. Using (134), we then obtain

$$\frac{M_p c^2}{\sqrt{1 - v_2^2/c^2}} = \frac{E_p}{\sqrt{f_{\text{out}}}}. \quad (138)$$

The same operations with (65), (28), and (136) yield

$$\frac{M_p c^2}{\sqrt{1 - v_2'^2/c^2}} = \frac{E_p'}{\sqrt{f_{\text{in}}}}. \quad (139)$$

In these formulas, as previously, all functions of $r$ are taken at the point of intersection $r = r^*$. Since the jump in energy is known [relation (137)], the jump in the velocity of the test shell can be determined from (138) and (139).

## 7. MASS EJECTION FROM A STAR CLUSTER

The dynamical processes near supermassive black holes (SBHs), quasars, blazars, and active galactic nuclei are characterized by violent events that give rise to jets and ejections. The formation of jets is commonly associated with processes that take place in magnetized accretion disks [18, 19]. The formation of quasi-spherical ejections, which are possibly observed in broad absorption lines, may prove to be related to other ejection mechanisms. In this section, based on the ballistic interaction between gravitating shells described in the preceding sections, we point out the possibility of shell ejection from the neighborhood of a SBH surrounded by a dense massive star cluster.

Numerical calculations for the collapse of a star cluster in the shell approximation [4, 9, 10] showed that, even if all shells were initially bound, after several intersections, some of the shells acquire enough energy to become unbound and to fly away to infinity. The remnant can be a stationary star cluster in the Newtonian approximation and a SBH in general relativity.

Ejections can be produced by the interaction between shells moving near a SBH. In a homogeneous star cluster with or without a SBH, stars evaporate through pair collisions with modest kinetic energy transfer. The formation of rapidly escaping stars is approximately a factor of 100 less probable, because collisions with weak momentum transfer prevail [20]. If the cluster is denser and contains several compact parts, then the collisions between these parts will be completely different; significant momentum transfer during a collision becomes possible. In this case, the gravitational interaction between compact parts can lead to high-velocity ejections, and if such an intersection takes place near a SBH, then the shell escape velocity from the cluster can account for an appreciable fraction of the speed of light $c$. Such a situation can arise through the collision of galaxies during a close encounter of their nuclei. In that case, one nucleus can pull part of the matter from the other nucleus in the form of collapsing shells. The interaction of such shells with cluster stars can lead not only to collapse onto the SBH but also to the reverse phenomenon: shell ejection with a velocity much higher than the fall velocity at a given radius; the shell will not fall to the SBH because of the large angular momentum of its stars.

The ejection mechanism manifests itself even for the interaction between two shells in the Newtonian approximation considered in Section 4. If two gravitationally bound shells with energies $\mathscr{E}_1 < 0$ and $\mathscr{E}_2 < 0$ that obey the equations of motion (93) and (94) intersect at point $r = r_1^*$, then their next intersection can occur at point $r = r_2^*$ farther from the center, i.e., at $r_2^* > r_1^*$. According to (90) and (91), the shell energies will be

$$\mathscr{E}_1' = \mathscr{E}_1 - \frac{km_1m_2}{r_1^*}, \quad \mathscr{E}_2' = \mathscr{E}_2 + \frac{km_1m_2}{r_1^*} \quad (140)$$

after the first intersection and

$$\mathscr{E}_1'' = \mathscr{E}_1' + \frac{km_1m_2}{r_2^*} = \mathscr{E}_1 - km_1m_2\left(\frac{1}{r_1^*} - \frac{1}{r_2^*}\right),$$
$$\mathscr{E}_2'' = \mathscr{E}_2' - \frac{km_1m_2}{r_2^*} = \mathscr{E}_2 + km_1m_2\left(\frac{1}{r_1^*} - \frac{1}{r_2^*}\right) \quad (141)$$

after the second intersection. If the absolute values of $\mathscr{E}_1$ and $\mathscr{E}_2$ are sufficiently small and if $r_1^*$ is moderately large and not too close to $r_2^*$, then $\mathscr{E}_2'' > 0$ and (outer) shell 2 after the intersection can go to infinity. Clearly, there is a broad class of such solutions, and one specific example (with the highest possible ejection velocity) is given in [8].

Naturally, this effect also remains in the relativistic theory of gravitation. If two gravitationally bound shells with energies $E_1 < m_1c^2$ and $E_2 < m_2c^2$ that move according to Eqs. (30)–(32) intersect at point $r = r_1^*$, then the energy transfer is described by formulas (82) and (84), (85):

$$E_1' = E_1 - \frac{kM_1(r_1^*)M_2(r_1^*)}{r_1^*}(-Q),$$
$$E_2' = E_2 + \frac{kM_1(r_1^*)M_2(r_1^*)}{r_1^*}(-Q). \quad (142)$$

Here, $Q$ is given by expression (70), in which all functions of $r$ are taken at point $r_1^*$. For simplicity, let us consider only those cases where the second intersection occurs at $r_2^* > r_1^*$ but at such a large $r_2^*$ that the Newtonian approximation can be used for estimates in this region. Thus, the shell energies after the second intersection will be

$$E_1'' = E_1' + \frac{km_1m_2}{r_2^*}$$
$$= E_1 - \left[\frac{kM_1(r_1^*)M_2(r_1^*)}{r_1^*}(-Q) - \frac{km_1m_2}{r_2^*}\right],$$
$$E_2'' = E_2' - \frac{km_1m_2}{r_2^*} \quad (143)$$
$$= E_2 + \left[\frac{kM_1(r_1^*)M_2(r_1^*)}{r_1^*}(-Q) - \frac{km_1m_2}{r_2^*}\right].$$

Now, an important circumstance is that, whatever the value of $r_1^*$, the first term in the square brackets in (143) satisfies the inequality

$$\frac{kM_1(r_1^*)M_2(r_1^*)}{r_1^*}(-Q) > \frac{km_1m_2}{r_1^*}. \quad (144)$$

This follows from the fact that $M_1(r) > m_1$ and $M_2(r) > m_2$ at any $r$ and, in addition, the absolute value of $Q$ is always larger than unity (see Footnote 3). Comparison of expressions (143), (144), and (141) indicates that the shell ejection effects in the relativistic region not only remain but can even be more intense.

## 8. A NUMERICAL REALIZATION OF THE EXACT SOLUTION

Let us now consider a numerical solution of the exact equations of motion for two intersecting shells. To calculate the motions of the shells between their intersections, we used Eqs. (29)–(36), where $m_{in}$, $m_{12}$, $m_{out}$, $m_1$, $m_2$, $L_1$, and $L_2$ are the free initial parameters of the system. It is also required to specify the initial shell radii; the calculation start time may be taken to be zero. We deduce expressions for the derivatives of the proper times $\tau_1$ and $\tau_2$ with respect to $t$ from (33)–(36). Substituting them into Eqs. (30) and (31) yields the equations of motion for the shells in time relative to an infinitely distant observer.

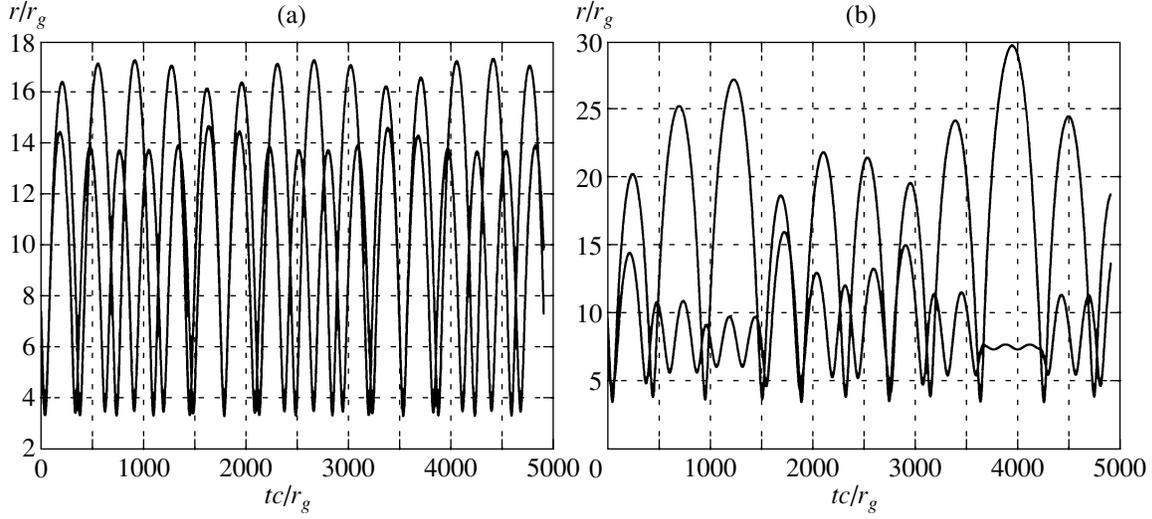

**Fig. 2.** The motion of two gravitationally bound shells with intersections when the mass of each shell accounts for 1% of the mass of the central body: (a) $m_1 = m_2 = 0.01 m_{in}$ and (b) $0.15 m_{in}$.

The energy of the shells is redistributed during their intersection, which can be taken into account by using Eq. (78). The shells were renamed after their intersection, which allows us to count as many intersections as we wish without being concerned about the shell numbers (shell 2 is always the outer one).

Figure 2 shows the motion of two shells around a central body. The rest masses of the shells were assumed to be identical and equal to 1 and 15% of the mass of the central body, and the initial relative energies $E_i/m_i c^2$ were 0.975 in both cases. Shell "beating" and energy transfer from one shell to the other, which manifests itself in changes in the radius of maximum shell recession from the center, are clearly seen in Fig. 2a. However, at larger shell masses in Fig. 2b, the shell energies and trajectories even after one intersection change so greatly that the next intersection can occur for quite different shell radii and velocity directions (i.e., the change in phase is comparable to the shell oscillation period itself). We clearly see chaotization of the shell motion from this example. Chaotic motions for various shell parameters in the Newtonian case are illustrated in [8].

To achieve the largest gain in energy of the escaping shell, by analogy with the Newtonian case, we should choose the shell parameters as follows: first, the initial total energies must be close to the rest energies $m_1 c^2$ and $m_2 c^2$; and, second, the first intersection must occur at a point as close to the gravitating center as possible, while the second intersection must occur as far as possible from this center. The characteristic relativistic potential can be used to satisfy these conditions. Near the peak of the potential curve (near $2r_g$), one of the shells can be arbitrarily long; the shell, as it were, sticks to the radius of the potential peak, which gives time for the other shell to fly far away (see Fig. 3).

Figure 4 shows the motion of shells with intersection and with the ejection of one shell after the second intersection. The rest masses of the shells were assumed to be identical and equal to 1% of the mass of the central body, $m_1 = m_2 = 0.01 m_{in}$ (Fig. 4a). The other initial parameters were taken to be $r_1 = 7.5$, $r_2 = 7.75$, $L_1 = 2.013$, $L_2 = 2.0279481$, $m_{out} - m_{12} = (1 - 10^{-12}) m_2$, and $m_{12} - m_{in} = (1 - 10^{-12}) m_1$. Here, $r_i$ and $L_i$ are given in units of $r_g$ and $m_i r_g c$, respectively, with $r_g = 2k m_{in}/c^2$. The first and second intersections occur at $r_1^* = 2.126104$ and $r_2^* = 43.8996$, respectively. The escaping shell acquires an energy $\Delta m c^2$ approximately equal to its kinetic energy $mv^2/2 = \Delta m c^2$, $\Delta m = 4.3604 \times 10^{-5} m_{in} \approx 4.4 \times 10^{-3} m_1$, which corresponds to the velocity at infin-ity $v = 0.0931c$. The initial and resulting parameters for the various calculations shown in Figs. 4a–4d and Fig. 5 are given in the table. Figures 6–8 show plots for variations in the radii of the first and second intersections and in the escape velocity of one of the shells to infinity for the conditions most favorable for shell ejection.

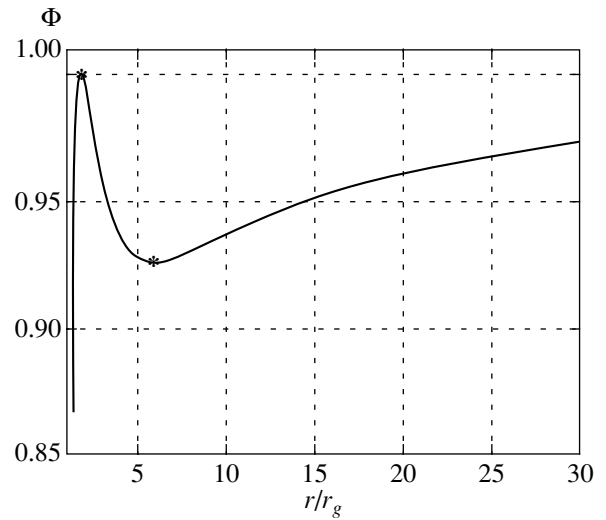

**Fig. 3.** The total (gravitational plus centrifugal) effective potential $\Phi$ for the motion of one shell, $r_g = 2k m_{in}/c^2$.

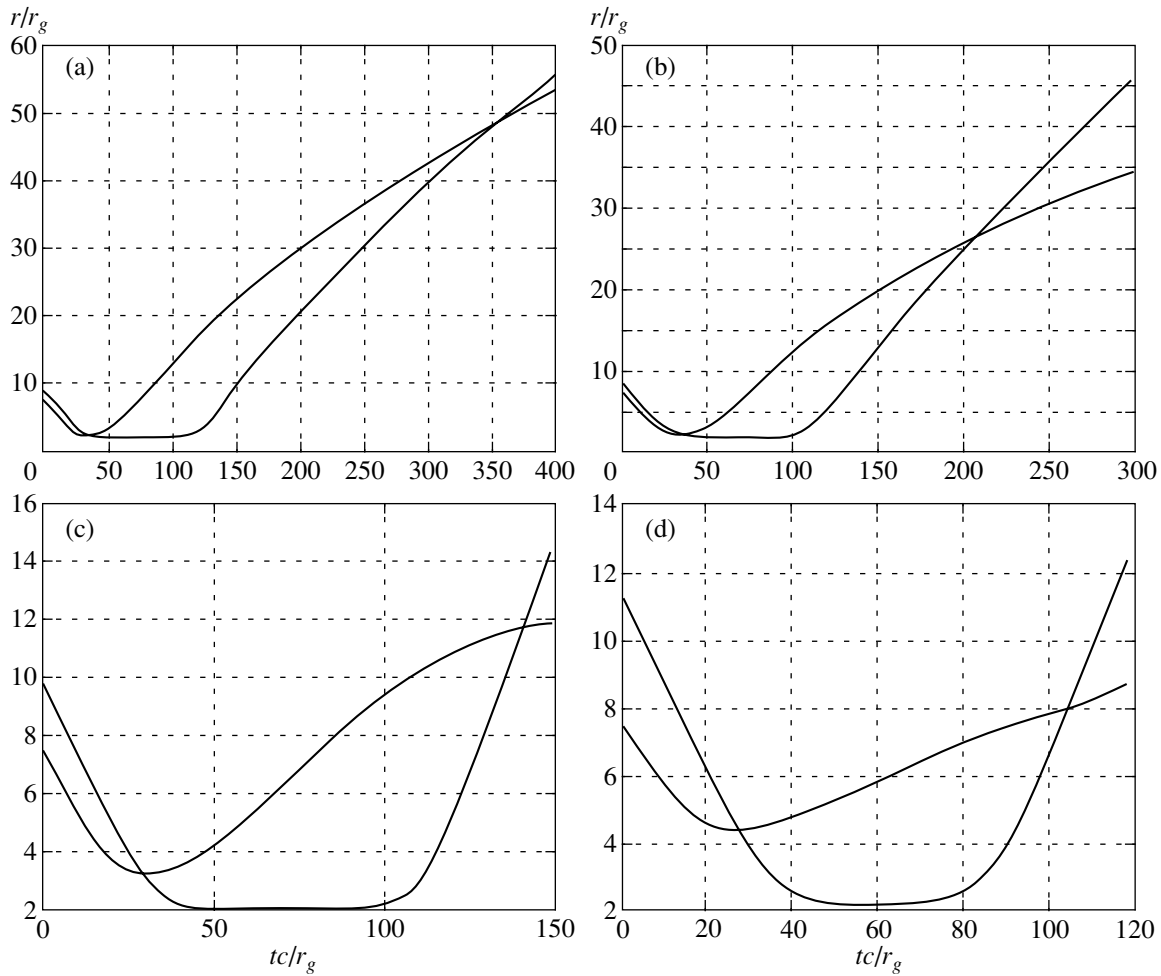

**Fig. 4.** The ejection of one of the shells after two intersections under the conditions most favorable for the attainment of the maximum ejection velocity: (a) $m_1 = m_2 = 0.01 m_{in}$, (b) $0.03 m_{in}$, (c) $0.15 m_{in}$, and (d) $0.30 m_{in}$.

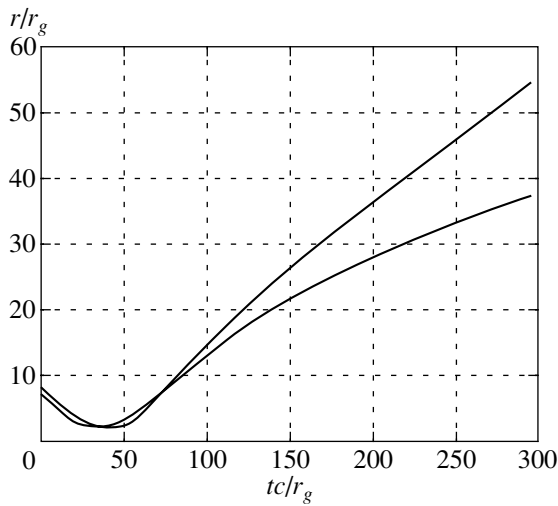

**Fig. 5.** The ejection of one of the shells after two intersections for initial parameters identical to those in Fig. 4b with the angular momentum of the particles constituting the second shell increased by 0.5%, which violates the most favorable conditions for ejection.

Figure 5 shows the motion of the same shells as those in Fig. 4b but with the angular momentum of shell 2 increased by half a percent. The latter causes the "sticking" phase to disappear; as a result, the second intersection occurs much earlier and the efficiency of the mechanism decreases sharply. The change in energy was found to be $\Delta m = 2.634 \times 10^{-4} m_{in}$, which accounts for about 0.88% of the shell rest mass; i.e., the efficiency of the mechanism decreased by 17% (see the table).

In conclusion, note that, as the shell masses rise, the efficiency of the ballistic ejection mechanism initially increases. However, when the shell rest masses reach about 20% of the mass of the central body, the mini-mum possible radius of the first intersection increases and the maximum possible radius of the second intersection decreases because of the strong interaction between the shells, which causes the efficiency of this scenario to decrease (see Figs. 6–8). In turn, as was mentioned above, a small deviation of the parameters from the optimal position results in a significant deviation from the limiting shell escape velocity.

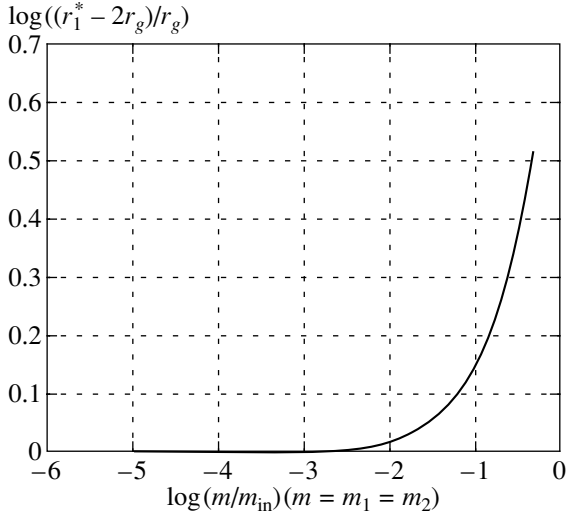

**Fig. 6.** $\log[(r_1^* - 2r_g)/r_g]$ versus logarithm of the mass ratio, $\log(m_1/m_{in})$; $r_1^*$ is the radius of the first intersection for equal rest masses of the shells, $m_1 = m_2$, under the conditions most favorable for ejection.

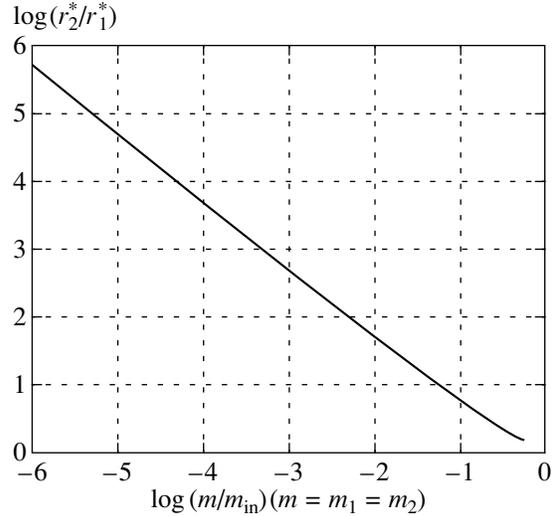

**Fig. 7.** $\log(r_2^*/r_1^*)$ versus $\log(m/m_{in})$; $r_2^*$ is the radius of the second intersection for the same conditions as in Fig. 6.

The calculations of the shell motion with simplified conditions during intersections from [9, 10] (see the Appendix) proved to be in good agreement with the exact calculations for low shell masses, $m_{1,2}/m_{in} \leq 0.03$. This is because, under the conditions most favorable for escape, the first intersection occurs near the point of minimum radius, where either $v_1 \ll c$ or $v_2 \ll c$ (the necessary condition for the validity of the simplified condition) and the second intersection occurs far from the center in the nonrelativistic region, where $v_1, v_2 \ll c$ as well. Thus, the conditions under which the approximate solution is in good agreement with the exact solution are satisfied.

For low-mass ($m_1, m_2 \ll m_{in}$) shells, we can take the radius of the first intersection to be $r_1^* = 2r_g$ and the radius of the second intersection to be infinite. As a result, the maximum energy carried away by a low-mass shell through the intersections with a shell of the same rest mass is proportional to the shell mass and the escape velocity is proportional to the square root of the shell mass. Using the first row in the table, we have

$$\frac{\Delta m}{m} \approx 0.044 \sqrt{\frac{m}{m_{in}}}, \quad \frac{v}{c} \approx 0.931 \sqrt{\frac{m}{m_{in}}}$$

for $m/m_{in} \ll 1$, where $m = m_1 = m_2$.

**Table**

|  | Fig. 4a | Fig. 4b | Fig. 4c | Fig. 4d | Fig. 5 |
|---|---|---|---|---|---|
| $m_1/m_{in}$ | 0.01 | 0.03 | 0.015 | 0.30 | 0.03 |
| $m_2/m_{in}$ | 0.01 | 0.03 | 0.015 | 0.30 | 0.03 |
| $r_1/r_g$ | 7.5 | 7.5 | 7.5 | 7.5 | 7.5 |
| $r_2/r_g$ | 7.75 | 8.5 | 9.8 | 11.27 | 8.5 |
| $r_1^*/r_g$ | 2.126104 | 2.3638 | 3.249 | 4.3698 | 2.3765 |
| $r_2^*/r_g$ | 43.8996 | 26.986 | 11.7076 | 7.9374 | 7.9345 |
| $L_1/m_1 c r_g$ | 2.013 | 2.05 | 2.285 | 2.61 | 2.05 |
| $L_2/m_2 c r_g$ | 2.0279481 | 2.0753315 | 2.305431 | 2.5393 | 2.0857082 |
| $(m_{out} - m_{12})/m_2$ | $1-10^{-12}$ | $1-10^{-12}$ | $1-10^{-12}$ | $1-10^{-12}$ | $1-10^{-12}$ |
| $(m_{12} - m_{in})/m_1$ | $1-10^{-12}$ | $1-10^{-12}$ | $1-10^{-12}$ | $1-10^{-12}$ | $1-10^{-12}$ |
| $\Delta m/m_{in}$ | $4.3604 \times 10^{-5}$ | $3.1889 \times 10^{-4}$ | $4.22686 \times 10^{-3}$ | $7.52 \times 10^{-3}$ | $2.634 \times 10^{-4}$ |
| $v/c$ | 0.0931 | 0.1447 | 0.2336 | 0.2198 | 0.1384 |
| $\Delta m/m_2$, % | 0.44 | 1.06 | 2.8 | 2.5 | 0.88 |

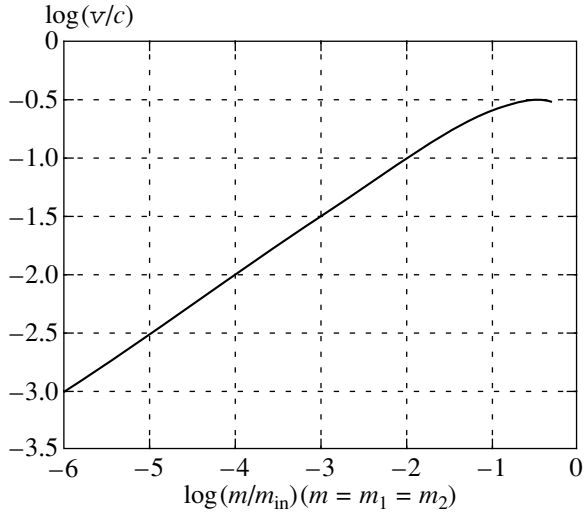

**Fig. 8.** Logarithm of the ejection velocity to infinity, $\log(v/c)$], versus $\log(m/m_{in})$ for the same conditions as in Fig. 6.

## ACKNOWLEDGMENTS

This study was supported in part by the Russian Foundation for Basic Research (project no. 99-02-18180), the INTAS-ECO (project no. 120), and the INTAS (project no. 00-491).

## APPENDIX

Since there was no exact solution to the problem of the intersection of relativistic gravitating shells in the course of the study by Bisnovatyi-Kogan and Yangurazova [9, 10], these authors proposed the following approximate continuity conditions at the point of intersection:

$$\frac{E_1}{\sqrt{1 - r_{g1}/r^*}} = \frac{E_1'}{\sqrt{1 - r_{g1}'/r^*}}, \quad \text{(A.1)}$$

$$E_1' + E_2' = E_1 + E_2. \quad \text{(A.2)}$$

Here, the total energies of the shells $E_1$, $E_2$, $E_1'$, and $E_2'$ are given by formulas (79) and (80); $r_{g1}$ and $r_{g1}'$ are[4]

$$r_{g1} = \frac{2k}{c^2}\left(m_{in} + \frac{E_1}{2c^2}\right),$$

$$r_{g1}' = \frac{2k}{c^2}\left(m_{in} + \frac{E_2'}{c^2} + \frac{E_1'}{2c^2}\right). \quad \text{(A.3)}$$

Equation (A.2) is the exact energy equation (81) and requires no discussion. In contrast, as our comparison with the exact theory shows (see below), Eq. (A.1) is approximate. It can be used only for low effective rest masses of the shells $M_1$ and $M_2$ (compared to the mass of the central body) together with the condition for the velocity of at least one of the shells being low (compared to the speed of light) at the intersection time. Our exact solution with the most favorable (for escape) conditions corresponds to the case where the first intersection occurs near the turning point (the minimum possible distance from the central body) of one of the shells, i.e., where the velocity of this shell is nearly zero. The energy transfer responsible for the ejection of one of the shells to infinity is determined by this first intersection, because the second intersection occurs far from the center, where the energy transfer is negligible. For these reasons, the results of our numerical calculations using conditions (A.1) and (A.2) for sufficiently low shell masses proved to be similar to those obtained by using the exact theory when describing the cases corresponding to maximum shell escape velocities.

Let us briefly explain the derivation of the approximate condition (A.1) from the exact solution. Consider the intersection of shells described in Section 3 for low (compared to $m_{in}c^2$) energies $E_1$ and $E_2$. In addition, we assume that the intersection does not occur too close to the gravitational radius of the central body and that, although the shell velocities at the intersection time can account for a sizeable fraction of the speed of light, they are, nevertheless, not ultrarelativistic. This means that

$$E_1, E_2 \ll m_{in}c^2, \quad f_{in}(r^*) \sim 1,$$

$$\sqrt{1 - v_1^2/c^2}, \sqrt{1 - v_2^2/c^2} \sim 1. \quad \text{(A.4)}$$

Below, $v_1$ and $v_2$ are the velocities given by relations (59)–(61) but taken at point $r = r^*$. It is easy to show that, under conditions (A.4), the equations of motion (30) and (31), to a first approximation, yield

$$E_1 = \left(\frac{M_1 c^2 \sqrt{f_{in}}}{\sqrt{1 - v_1^2/c^2}}\right)_{r = r^*},$$

$$E_2 = \left(\frac{M_2 c^2 \sqrt{f_{in}}}{\sqrt{1 - v_2^2/c^2}}\right)_{r = r^*}, \quad \text{(A.5)}$$

---

[4] Actually, the terms $E_1/2c^2$ and $E_1'/2c^2$ related to shell self-gravitation were taken in [9, 10] without the factor 1/2. The more accurate expressions (A.3) were used later.

whence it also follows that conditions (A.4) mean $M_1$, $M_2 \ll m_{\text{in}}$. In this approximation, formula (85) for energy transfer can be written as

$$\Delta E = \frac{k(1 - v_1 v_2/c^2)}{c^4 r^* f_{\text{in}}(r^*)} E_1 E_2. \quad (A.6)$$

If we also add the requirement that the intersection occur near the turning point of one of the shells, i.e., for $v_1 v_2/c^2 \ll 1$, then it follows from (A.6) that

$$\Delta E = \frac{k E_1 E_2}{c^4 r^* f_{\text{in}}(r^*)}. \quad (A.7)$$

It is easy to show that the same expression for $\Delta E$ also follows from Eqs. (A.1) and (A.2) if we express $E_1 - E_1' = \Delta E$ in them as a function of $E_1$, $E_2$, and $m_{\text{in}}$ and take the first term of its expansion in small parameters $E_1/m_{\text{in}}c^2$ and $E_2/m_{\text{in}}c^2$ for $f_{\text{in}}(r^*) \sim 1$.

## REFERENCES

1. M. Hénon, Ann. Astrophys. **27**, 83 (1964).
2. M. Hénon, Astron. Astrophys. **24**, 229 (1973).
3. S. L. Shapiro and S. A. Teukolsky, Astrophys. J. **298**, 34 (1985).
4. L. R. Yangurazova and G. S. Bisnovatyi-Kogan, Astrophys. Space Sci. **100**, 319 (1984).
5. J. R. Gott, Astrophys. J. **201**, 296 (1975).
6. J. L. Spitzer and H. M. Hart, Astrophys. J. **164**, 399 (1971).
7. J. E. Chase, Nuovo Cimento B **67**, 136 (1970).
8. M. V. Barkov, V. A. Belinski, and G. S. Bisnovatyi-Kogan, astro-ph/0107051; Mon. Not. R. Astron. Soc. **334**, 338 (2002)
9. G. S. Bisnovatyi-Kogan and L. R. Yangurazova, Astrofizika **27**, 79 (1987).
10. G. S. Bisnovatyi-Kogan and L. R. Yangurazova, Astrophys. Space Sci. **147**, 121 (1988).
11. V. Berezin and M. Okhrimenko, Class. Quantum Grav. **18**, 2195 (2001).
12. A. Neronov, hep-th/0109090.
13. J. Khoury, B. Ovrut, P. Stainhardt, and N. Turok, hep-th/0103239.
14. M. Bucher, hep-th/0107148.
15. D. Langlois, K. Maeda, and D. Wands, gr-qc/0111013.
16. W. Israel, Nuovo Cimento B **44**, 1 (1966).
17. T. Dray and G.'t Hooft, Commun. Math. Phys. **99**, 613 (1985).
18. R. V. E. Lovelace, Nature **262**, 649 (1976).
19. G. S. Bisnovatyi-Kogan and S. I. Blinnikov, Pis'ma Astron. Zh. **2**, 489 (1976) [Sov. Astron. Lett. **2**, 191 (1976)].
20. V. A. Ambartsumyan, Uch. Zap. Leningr. Gos. Univ. **22**, 19 (1938).